\documentclass{pasa}%

\usepackage{graphicx}
\usepackage{gensymb} 
\usepackage{xfrac}

\newcommand{\red}[1] {#1}

\newcommand{\Ncoincnights}{14 }
\newcommand{\Ncoinctime}{366 }
\newcommand{\NBnights}{24 } 
\newcommand{\NBtime}{358.5 }

\newcommand{\fovdegree}{11909.35}
\newcommand{\fovreduction}{0.82} 
\newcommand{\fovdegreecorr}{9765.7} 

\newcommand{\sdlimittime}{129.25} 
 
\newcommand{\sdlimitepochs}{232645}

\newcommand{\surfacedensitylimit}{1.32 \times 10^{-9}} %

\title[Southern-Hemisphere all-sky radio transient monitor]{A Southern-Hemisphere all-sky radio transient monitor for SKA-Low prototype stations}

\author[M.~Sokolowski et al.]
{M.~Sokolowski$^{1}$\thanks{marcin.sokolowski@curtin.edu.au}, 
R.~B.~Wayth$^{1,2}$, 
N.~D.~R.~Bhat$^1$, 
D.~Price$^{1}$, 
J.~W.~Broderick$^{1}$, 
G.~Bernardi$^{3}$, 
P.~Bolli$^{3}$, 
R.~Chiello$^{4}$, 
G.~Comoretto$^{3}$, 
B.~Crosse$^1$, %
D.~B.~Davidson$^1$, %
G.~Macario$^{3}$, 
A.~Magro$^{5}$, 
A.~Mattana$^{6}$, 
D.~Minchin$^1$, 
A.~McPhail$^{7}$, 
J.~Monari$^{6}$, 
F.~Perini$^{6}$, 
G.~Pupillo$^{3}$, 
G.~Sleap$^{7}$, 
S.~Tingay$^1$, 
D.~Ung$^{1}$,
A.~Williams$^{7}$ 
\affil{$^1$International Centre for Radio Astronomy Research, Curtin University, Bentley, WA 6102, Australia}%
\affil{$^2$ARC Centre of Excellence for All Sky Astrophysics in 3 Dimensions (ASTRO 3D), Australia}
\affil{$^3$Osservatorio Astrofisico di Arcetri, Istituto Nazionale di Astrofisica, Florence, Italy}
\affil{$^4$University of Oxford, Denys Wilkinson Building, Oxford, UK}
\affil{$^5$Institute of Space Sciences and Astronomy, University of Malta, Msida, Malta}
\affil{$^6$Istituto di Radioastronomia, Istituto Nazionale di Astrofisica, Bologna, Italy}
\affil{$^7$Curtin Institute of Radio Astronomy, GPO Box U1987, Perth, WA 6845, Australia}
}%

\jid{PASA}
\doi{10.1017/pas.\the\year.xxx}
\jyear{\the\year}

\usepackage{aas_macros}
\usepackage{hyperref} 
\hypersetup{colorlinks,citecolor=blue,linkcolor=blue,urlcolor=blue}

\hypersetup{draft}

\begin{document}

\begin{frontmatter}
\maketitle

\begin{abstract}
We present the first southern-hemisphere all-sky imager and radio-transient monitoring system implemented on two prototype stations of the low-frequency component of the Square Kilometre Array (SKA-Low). Since its deployment the system has been used for real-time monitoring of the recorded commissioning data. Additionally, a transient searching algorithm has been executed on the resulting all-sky images. It uses a difference imaging technique to enable identification of a wide variety of transient classes, ranging from human-made radio-frequency interference to genuine astrophysical events.
Observations at the frequency 159.375\,MHz and higher in a single coarse channel ($\approx$0.926\,MHz) were made with 2\,s time resolution, and multiple nights were analysed generating thousands of images. Despite having modest sensitivity ($\sim$ few Jy/beam), using a single coarse channel and 2-s imaging, the system was able to detect multiple bright transients from PSR B0950+08, proving that it can be used to detect bright transients of an astrophysical origin. 
The unusual, extreme activity of the pulsar PSR B0950+08 (maximum flux density $\sim$155\,Jy/beam) was initially detected in a ``blind'' search in the 2020-04-10/11 data and later assigned to this specific pulsar. The limitations of our data, however, prevent use from making firm conclusions of the effect being due to a combination of refractive and diffractive scintillation or intrinsic emission mechanisms. 
The system can routinely collect data over many days without interruptions; the large amount of recorded data at 159.375 and 229.6875 MHz allowed us to determine a preliminary transient surface density upper limit of $\surfacedensitylimit \text{deg}^{-2}$ for a timescale and limiting flux density of 2\,s and 42\,Jy, respectively. In the future, we plan to extend the observing bandwidth to tens of MHz and improve time resolution to tens of milliseconds in order to increase the sensitivity and enable detections of Fast Radio Bursts below 300\,MHz.

\end{abstract}

\end{frontmatter}

\section{INTRODUCTION}
\label{sec:intro}

All-sky imaging is a very powerful and unique feature of low-frequency interferometers operating below 400\,MHz, where the individual antennas can see the entire hemisphere. Several all-sky monitoring systems have been implemented in the Northern Hemisphere.  
The Amsterdam-ASTRON  Radio Transient Facility And Analysis Centre \citep[AARTFAAC;][]{2016JAI.....541008P} is a parallel transient detection instrument operating as a subsystem of the LOw Frequency ARray \citep[LOFAR;][]{2013A&A...556A...2V} observing at frequencies between 10 and 90\,MHz. Similarly, the Long Wavelength Array \citep[LWA;][]{2013ITAP...61.2540E} observes in the $10-88$\,MHz frequency band. 

These facilities are used for monitoring large swaths of the sky for transients. They have reported multiple results related to astrophysical transients ranging from detections of local Jovian bursts \citep{2016ApJ...826..176I}, extremely bright pulses from pulsars such as PSR B0950+08 \citep{2020MNRAS.497..846K}, flare stars \citep{2020MNRAS.494.4848D}, meteor radio afterglows \citep{2014ApJ...788L..26O}, limits on prompt emission from Gamma-Ray Bursts \citep{2014ApJ...785...27O,2018ApJ...864...22A}, to detections of short-timescale transients that are of yet unknown origin \citep{2019ApJ...874..151V,2020arXiv200311138K,2020arXiv200313289K}. Furthermore, the all-sky imaging searches have resulted in very stringent limits on transient surface densities \citep[e.g.][]{2019ApJ...886..123A,2020arXiv200313289K}. 
Extremely interesting detections already obtained by all-sky monitoring systems prove that these systems are powerful transient-instruments complementing wide-field and all-sky telescopes using other electromagnetic wavelengths or messengers (e.g. neutrino telescopes, gravitational waves detectors etc.), and have a potential of generating high impact scientific results. Especially, in the light of the recent detections of Fast Radio Bursts (FRBs) below 400\,MHz, such as the low-freqeuncy detections of repeating FRB 20180916B discovered by Canadian Hydrogen Intensity Mapping Experiment \citep[CHIME/FRB;][]{2018ApJ...863...48C,2019ApJ...885L..24C} and recently detected by LOFAR at 110 - 188\,MHz \citep{2020arXiv201208372P,2020arXiv201208348P} and Sardinia Radio Telescope \citep{2020ApJ...896L..40P} or FRB\,20200125A discovered by Green Bank Telescope at 350 MHz \citep{2020ApJ...904...92P}. Although the Northern Hemisphere systems cover some fractions of the southern sky, to date, no dedicated system capable of continuous monitoring of the entire Southern Hemisphere has existed.

We present the first all-sky transient monitoring facility in the Southern Hemisphere realised on the prototype stations of the low frequency component of the Square Kilometre Array \cite[SKA-Low;][]{2009IEEEP..97.1482D}\footnote{www.skatelescope.org}. This system takes advantage of the two prototype stations, the Engineering Development Array 2 (EDA2; Wayth et al., in preparation) and Aperture Array Verification System 2 \citep[AAVS2;][]{SPIE_Andre_van_Es,DavidsonBolli_etal_2020}, which were deployed at the MRO in 2019 to verify the technology and performance of different antenna designs against the SKA-Low requirements. These stations can observe in the same frequency band (50 - 350\,MHz) as intended for the full SKA-Low. More importantly to the presented project, they can be operated as standalone interferometers and form all-sky images from correlation products (visibilities) of all antenna pairs within the station. These images can then be searched for transients either in real-time or off-line. 

Real-time all-sky imaging has recently been implemented, and multiple long commissioning observations were performed with one or both stations observing in parallel in the same or different frequency bands. Although the most common transient candidates are due to radio-frequency interference (RFI, due to transmissions or reflections from aircraft, satellites or meteors), similar to those reported by \citet{2020PASA...37...39T} in the FM frequency band (98.44\,MHz), several transients of confirmed astrophysical origin have also been identified. The brightest and most interesting amongst them were extremely bright transients ($\sim$150\,Jy/beam, i.e. fluence $\sim$300 kJy/beam\,ms) from the pulsar PSR B0950+08. The pulsar PSR B0950+08 is a relatively long-known pulsar first reported as Cambridge Pulsed source CP0950 by \citep{1968Natur.218..126P}. Nevertheless, it has recently been generating a lot of attention due its exceptional brightness; its high variability; the detections of its extremely bright events by \citet{2020MNRAS.497..846K} (potentially similar to the giant pulses produced by the Crab pulsar); the recent confirmation of the surrounding Pulsar Wind Nebula \citep[PWN;][]{2020MNRAS.495.2125R}; and brightness enabling studies of the substructure of individual pulses in high-time resolution \citep{2020PASA...37...34M}.

The presented transient detection system is at a very early development stage and many further improvements are planned, such as for example automatic classification capability to enable efficient excision of RFI and other non-astrophysical transients. However, the excision of RFI will be significantly improved with the planned increase in bandwidth (to $\sim$50\,MHz) and time resolution to tens of milliseconds. 

This paper is organised as follows. In Section~\ref{sec:skalow_stations} we present the SKA-Low stations which were used to develop the all-sky transient monitoring system. In Section~\ref{sec:observations} we describe the data acquisition mode and observations used in this paper. In Section~\ref{sec:realtime_allsky_imager} we present the all-sky imaging system and all stages of data pre-processing, calibration and imaging leading to all-sky images used as a basis for the transient monitoring. In Section~\ref{sec:realtime_transient_monitor} we present the all-sky transient monitoring system and the early version of transient identification and classification. In Section~\ref{sec:results} we discuss the preliminary results obtained with this system ranging from RFI-related events to transients from astrophysical objects, especially the extreme activity of the pulsar PSR B0950+08. In Section~\ref{subsec:frb_limits} we describe low-frequency upper limits on the flux density of two FRBs, and in Section~\ref{subsec:rate_limits} a preliminary upper limit on transient surface density is discussed. In Section~\ref{subsec:other_aplications} we outline potential other applications of the real-time imaging pipeline. Finally, a summary of this work is provided in Section~\ref{sec:summary} and in Section~\ref{sec:future} we discuss future improvements in the system.

\section{SKA-LOW PROTOTYPE STATIONS}
\label{sec:skalow_stations}

\begin{figure*}[h!]
    \includegraphics[width=\columnwidth,angle=0]{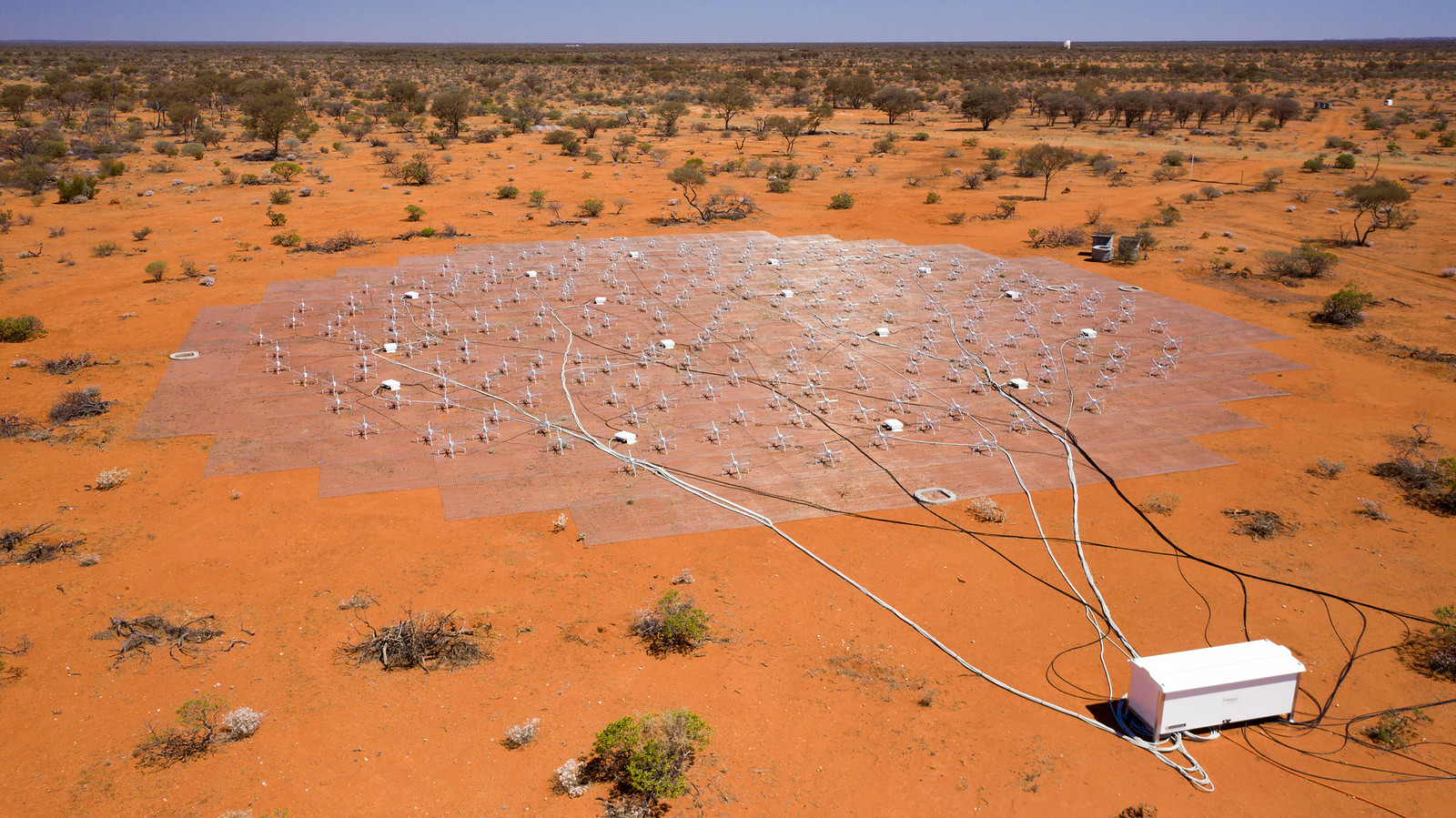} 
    \includegraphics[width=\columnwidth,angle=0]{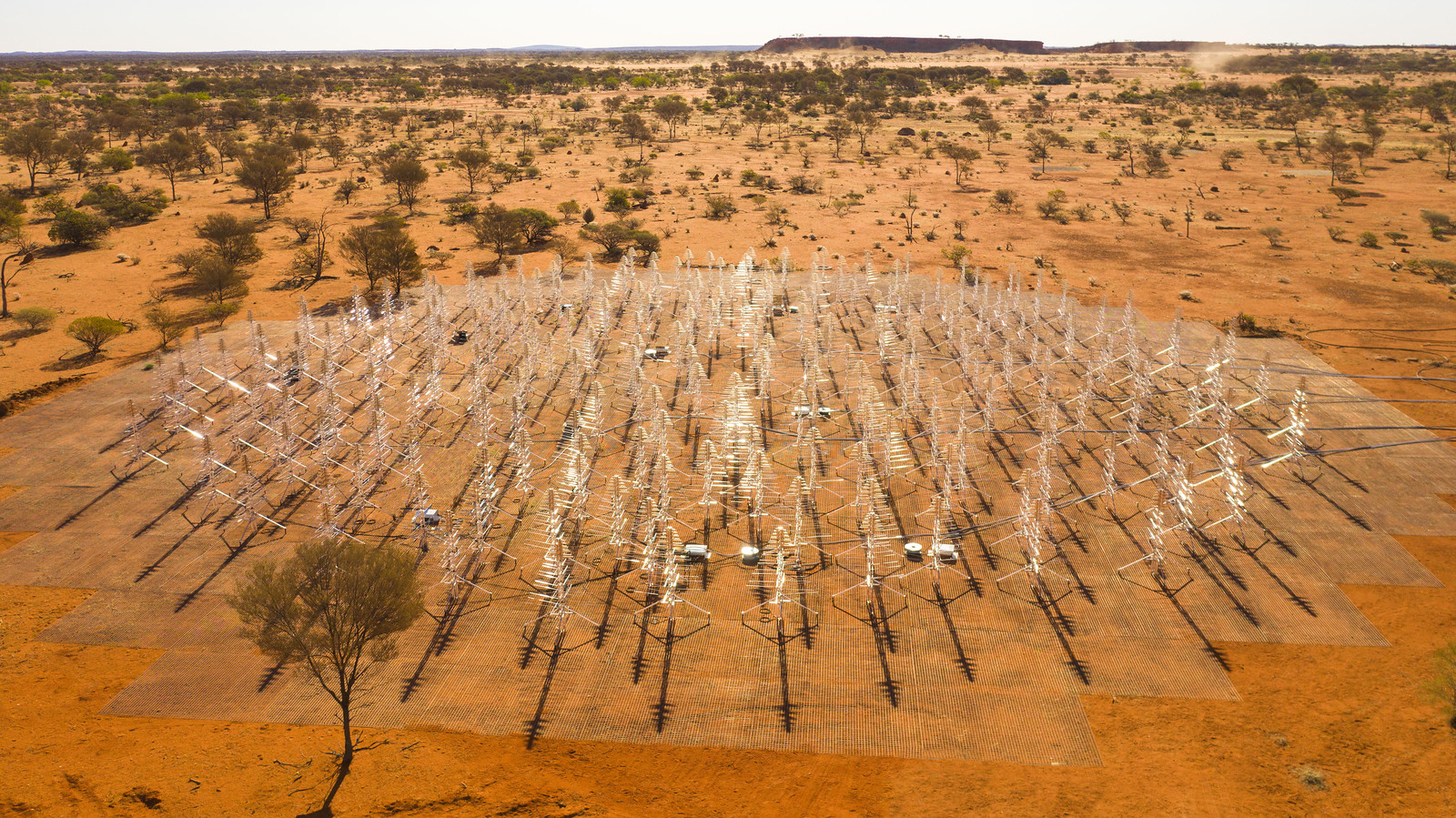} 
    \caption{An aerial view of the SKA-Low prototype stations EDA2 composed of 256 MWA bow-tie dipoles (left image) and AAVS2 composed of 256 SKALA4.1 antennas (right image), which were used for this paper.}
    \label{fig_eda2_and_aavs2_images}
\end{figure*}

The SKA-Low will be a low-frequency (50-350\,MHz) radio-telescope of an unprecedented collecting area and sensitivity. It will consist of 512 stations each composed of 256 dual-polarised antennas. 
In 2016, the first full-scale (256 dual-polarised antennas) prototype SKA-Low station, the Engineering Development Array 1 \citep[EDA1;][]{2017PASA...34...34W} was deployed at the MRO. It was composed of 256 dual-polarisation bow-tie dipoles of the same design as used in the Murchison Widefield Array \citep[MWA;][]{2013PASA...30....7T,2018PASA...35...33W} arranged in pseudo-random SKA-Low layout within a 35\,m diameter. It was deployed in order to assess applicability of the MWA-like technology (bow-tie dipoles and analogue beamforming) for the SKA-Low and as a reference for the Aperture Array Verification System 1 (AAVS1; Bentham et al. submitted), which was deployed at the MRO in 2017.

In 2019, based on these experiences, two further prototype stations were deployed at the MRO. They use the same signal chain technology and antenna layout as their predecessor AAVS1 station, but the antenna designs in both stations are different. The left panel of Figure~\ref{fig_eda2_and_aavs2_images} shows the EDA2 station, which as its predecessor (EDA1) consists of 256 MWA bow-tie dipoles, whilst the AAVS2 station shown in the right panel of Figure~\ref{fig_eda2_and_aavs2_images} is composed the SKALA4.1 antennas \citep{Bolli_etal2020,8520395}. The diameter of the AAVS2 station (maximum distance between antenna centres $\approx$38\,m) has been increased by about 10\% with respect to the other stations (EDA1, EDA2 and AAVS1).

The analogue signals from individual antennas (X and Y polarisations) are converted to optical signals near the antennas and transported over the 5.5\,km fiber to the central processing facility. The EDA2 and AAVS2 both use Tile Processing Modules \citep[TPM;][]{2017JAI.....641014N} to digitise incoming signals. The TPM is a 32-input signal processing board designed for SKA-Low. Each TPM digitises 32 inputs at 800 Msamples/s with 8-bit resolution, for a total ingest of 25.6 GB/s per board (409.6 GB/s per station). 

Steam-processing firmware running on the TPM boards coarsely channelises the incoming voltage streams into 512 channels of width $\approx$0.926 MHz; this firmware is detailed in \citep{2017JAI.....641015C}. The EDA2 and AAVS2 are connected by high-speed Ethernet to a software correlator running on commercial-off-the-shelf compute hardware with both stations using exactly the same rack mounted Dell servers each with two Intel Xeon Gold 6226 2.7 GHz, 192 GB RAM, 64 TB of SSD hard-drives in RAID5 for the data, two 240 GB solid-state drives in RAID1 for the operating system, NVIDIA Tesla V100 16GB GPU and one Mellanox ConnectX-5 dual port 40/100 Gb Ethernet card. We use a custom Ethernet packet capture code and the \textsc{xGPU} software correlator \citep{xGPU} to perform cross-correlation with all 256 inputs in real time. 

Thus, both stations form 256 element, dual-polarisation interferometers that can be used to produce all-sky images with standard interferometric calibration and imaging techniques. The TPMs can also generate real-time station beams to be correlated with other stations, which will be a typical operation mode for the SKA-Low. However, we do not compute inter-station cross correlations for the work presented here. Nevertheless, this functionality may also be used by the presented system to automatically form station beams in the direction of transients identified by the real-time system or provided by external alerts.

\section{OBSERVATIONS}
\label{sec:observations}

Multiple long (typically longer than 24\,hours) observations have been performed at several frequencies since both stations were fully deployed at the MRO in 2019. A lot of these observations were conducted with both stations collecting correlated data simultaneously at 2\,s resolution and single coarse channel of $\approx$0.926\,MHz bandwidth.
Presently, the system can collect correlated data in a single coarse channel, which is its main limitation and will be discussed in Section~\ref{sec:future}. 
 Many nights of observations dedicated to development of the presented system have also been collected and Table~\ref{tab_candidates} contains a summary of the observations used for this paper. All the presented data were collected at frequencies equal or above 159.375\,MHz and stations recorded either the same frequency channel or AAVS2 observed at a higher frequency channel than the EDA2.

\section{REAL-TIME ALL-SKY IMAGER}
\label{sec:realtime_allsky_imager}

\begin{figure*}[h!]

    \includegraphics[width=\textwidth]{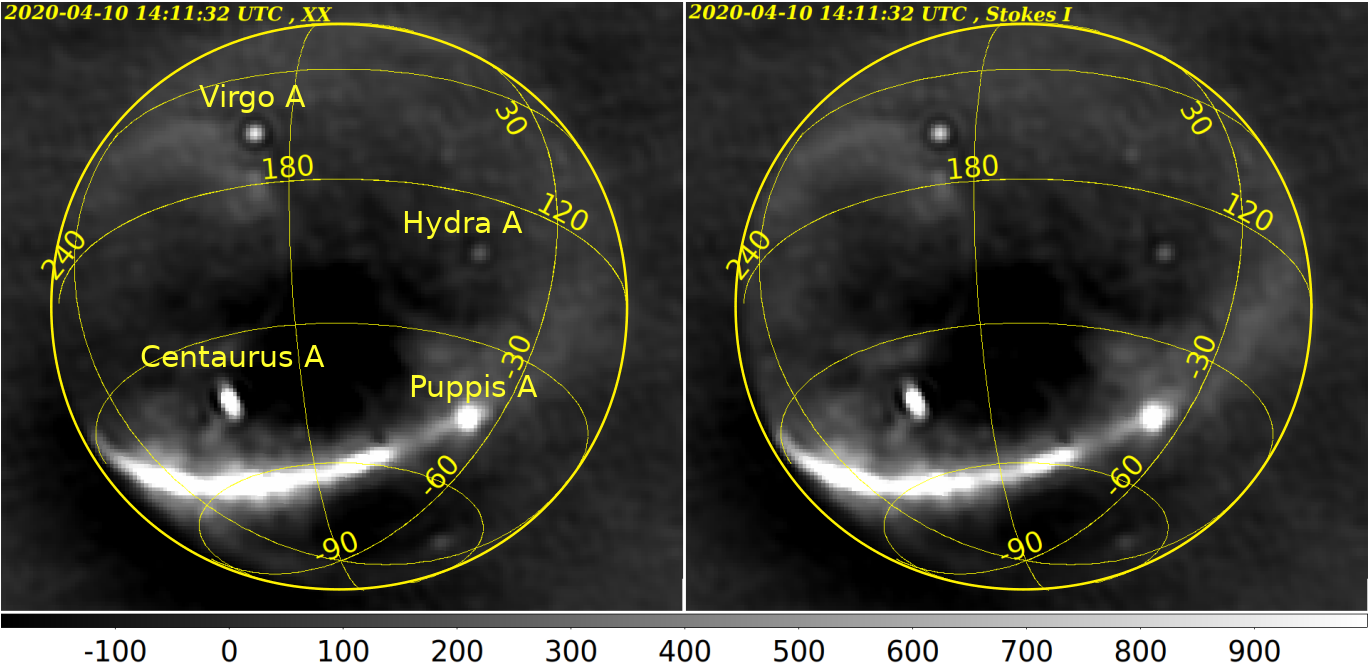}
    \caption{Examples of 2\,s all-sky images from the EDA2 at 159.375\,MHz collected on 2020-04-10 at 14:11:32 UTC. Left: XX image (YY image is virtually the same and not shown). \red{The brightest sources are labelled.} Right: Stokes I image, i.e. average of the beam corrected XX and YY images. The corresponding images from the AAVS2 station are very similar and not shown for brevity.}
    \label{fig_allsky_image_example}
\end{figure*}

\subsection{DATA PRE-PROCESSING}
\label{subsec:data_pre_processing}

The correlation products in $\approx$2\,s resolution are saved in \textsc{HDF5} format\footnote{https://www.hdfgroup.org/}, which is envisaged as the data format for the SKA telescope. The system waits for the new HDF5 file to be collected and converts it into the UV FITS file \citep{uvfits} in the native time resolution using the current metadata information, which includes list of flagged antennas, antenna positions etc. \red{We used auto-correlation spectra to identify and flag antennas with very low power, and antennas which did not calibrate well (i.e. calibration solutions as a function of frequency did not have clear linear form).
}
In the conversion process, the zenith is used as the phase center, and therefore the resulting images are zenith-centered (Sec.~\ref{subsec:allsky_imaging}). Time averaging is available in the correlator, but the system was always running at the currently highest possible time resolution of 1.96\,s.
Further processing was performed with the \textsc{Miriad} data processing suite \citep{1995ASPC...77..433S}. In the next step pre-calculated calibration solutions are applied to the data in UV format (Sec.~\ref{subsec:calibration}). 

\subsection{CALIBRATION}
\label{subsec:calibration}

The full band calibration scheme for the stations is an extension of the procedure described by Bentham et al. (submitted) \red{ and it will be briefly summarised here. Due to current bandwidth limitations (only one coarse channel $\approx$0.93\,MHz can be collected at a time), the full band calibration observations are performed as a loop (so called ``calibration loop'') over all 512 frequency channels and only short (2\,s) snapshots of correlated data are recorded by both stations, which takes about 30\,min to complete.
Therefore, longer calibration snapshot are not practicable because calibration data acquisition would take too long time, and change of source's position within the dipole beam would further complicate the calibration procedure. Due to this limitations we used transiting Sun as a phase and flux calibrator, which gives signal to noise ratio (SNR) $\sim$17000 in 2\,s images.}

\red{The ``calibration loop'' has been performed every few days, and short (2\,s) snapshots of correlated data have been recorded by both stations in each frequency channel around midday.
Using the Sun as a calibrator is justified at the frequencies of interest where the Sun is a dominant and unresolved radio source.
Further, over the last few years the solar activity cycle 25 has been at its minimum and it allowed us to also use the Sun a reliable flux calibrator.}

The \textsc{Miriad} task \textsc{mfcal}, with the quiet Sun flux model \citep{2009LanB...4B..103B}, was used to compute calibration solutions \red{(both phase and amplitude)}. Projected baselines shorter than $5\lambda$ (where $\lambda$ is the observing wavelength) were excluded to minimise the contribution from Galactic extended emission. The resulting calibration solutions are saved and the set of latest calibration solutions is updated on the data acquisition computer. These latest calibration solutions are automatically picked up by the real-time imaging pipeline when it is started. 

\subsubsection{PHASE CALIBRATION}
\label{subsec:phase_calibration}

\red{In the early commissioning stage the calibration loop was executed to calculate initial phase calibration solutions over the entire band and a linear function fitted to the resulting phase vs. frequency dependence. This fit yielded unaccounted time delays (in the range from -50 to +80\,ns) for every antenna, which were incorporated into the station configuration files and are always uploaded to TPM firmware every time station is initialised (before every any new observations are performed). Therefore, presently the unaccounted delays for all antennas are nearly zero, and when calibration loop is executed the resulting phase as a function of frequency dependence is almost a horizontal line for for each antenna. }

\red{It was confirmed during the early commissioning stage that phase calibration solutions are stable over long periods of time.
Moreover, stable phase behavior was confirmed with nearly 18 months of monitoring.
Using 24 calibrations in April 2020, we verified that when the time delays uploaded to firmware at the station initialisation step are used (nearly unchanged for 18 months or so), the mean (of 24 calibrations) antenna delays calculated by the calibration procedure were within 1\,ns (i.e. $\sim$60\degree\, at 160\,MHz) and standard deviation of additional delays fitted to phase vs. frequency by the calibration loop was very small, $\sim$0.084\,ns (corresponding to standard deviation of phase $\sim$5\degree\,at 160\,MHz). Therefore, given that when stations are initialised with the initial delay values resulting in phase errors below 60\degree\, it is possible to obtain good quality images even without applying any additional phase calibration. Nevertheless, we execute calibration loop every couple of days and the resulting phase calibration solutions (i.e. corrections with respect to the delays in the TPMs) are applied to the data before imaging, which corrects for the residual unaccounted delays.}

\subsubsection{FLUX CALIBRATION}
\label{subsec:flux_calibration}

\red{For accurate flux calibration, the apparent flux density of the Sun was calculated by multiplying the flux density predicted by the \citet{2009LanB...4B..103B} model by the response of the dipole beam pattern \citep{9232307} in the direction of the Sun. These beam patterns were simulated in FEKO electromagnetic simulation software. It was found during the station sensitivity studies that applying a single calibration (from the transiting Sun) to long observations may result in flux density errors of the order of 20 -- 30 \%. Similar variations were identified in the amplitudes of calibration solutions over many hours of calibration using low-frequency all-sky sky models, such as the sky image at 408\,MHz \citep{1982A&AS...47....1H}, the so called ``Haslam map'', scaled down to low frequencies using a spectral index of $-2.55$ \citep{spectral_indexASU2019} or Global Sky Model \citep{2008MNRAS.388..247D}. These variations of amplitudes of calibration solutions have been found to be mainly due to diurnal changes in ambient temperature. We have also tested flux density vs. time (lightcurves) of two bright calibrators (Hydra A and Virgo A), and found that their flux variations were within 10\% when they were above elevation 40\degree\, and 50\degree\, for EDA2 and AAVS2 stations respectively. The inaccuracy of flux density measurements at lower elevations stem mainly from the inaccuracy using a single dipole beam pattern for all station antennas, which may be significantly different from Embedded Elements Patterns (EEPs)\footnote{Beam patterns of individual dipoles in the array.} especially for the AAVS2 station using more complex antenna and more affected by the mutual coupling effects \citep{9232307}.}

\subsubsection{OTHER CALIBRATION METHODS}
\label{subsec:other_calibration_methods}

\red{Although the routine calibration procedure uses Sun as a phase and flux calibrator other calibration methods have also been successfully tested. The earlier mentioned calibration using low-frequency all-sky model (so called ``all-sky model calibration'') leads to similar calibration solutions and is the most likely replacement for the currently used procedure. Especially, that the minimum of solar cycle 25 comes to an end, and more active Sun may soon become a very inaccurate calibrator. We note that using individual bright calibrators (so called ``A-team sources''), such as Centaurus A, Hercules A, Hydra A, Pictor A, 3C444, Fornax A and Virgo A, is limited by SNR and sidelobes. These calibrators can, at best, provide SNR of the order of a few hundred (maybe $\sim$1000 for Centaurus A provided that good model of this source is used). However, this way of calibration have not been extensively tested yet, and is planned in the future. Especially, once the instantaneous bandwidth will be increased and/or longer calibration observations become practicable. We are expecting that the all-sky model (including A-team sources) calibration will be the most accurate method applicable over a wide range of local sidereal times (LSTs). Finally, novel calibration methods such as holography also yield very promising results (Kiefner et al. (submitted)), and may soon become an viable alternative to standard, visibilities based methods of station calibration.}

\subsection{ALL-SKY IMAGING}
\label{subsec:allsky_imaging}

The visibilities are calibrated using the set of the latest calibration solutions generated by the procedure described in Sections~\ref{subsec:data_pre_processing} and ~\ref{subsec:calibration}. The same set of calibration solutions is applied to all the data collected during a single acquisition as they are sufficiently stable (Sec.~\ref{subsec:calibration}). Therefore, it is not critical to dynamically update calibration solutions during the acquisitions, but this improvement is planned in the near future.

\begin{figure*}[h!]
    \includegraphics[width=\textwidth]{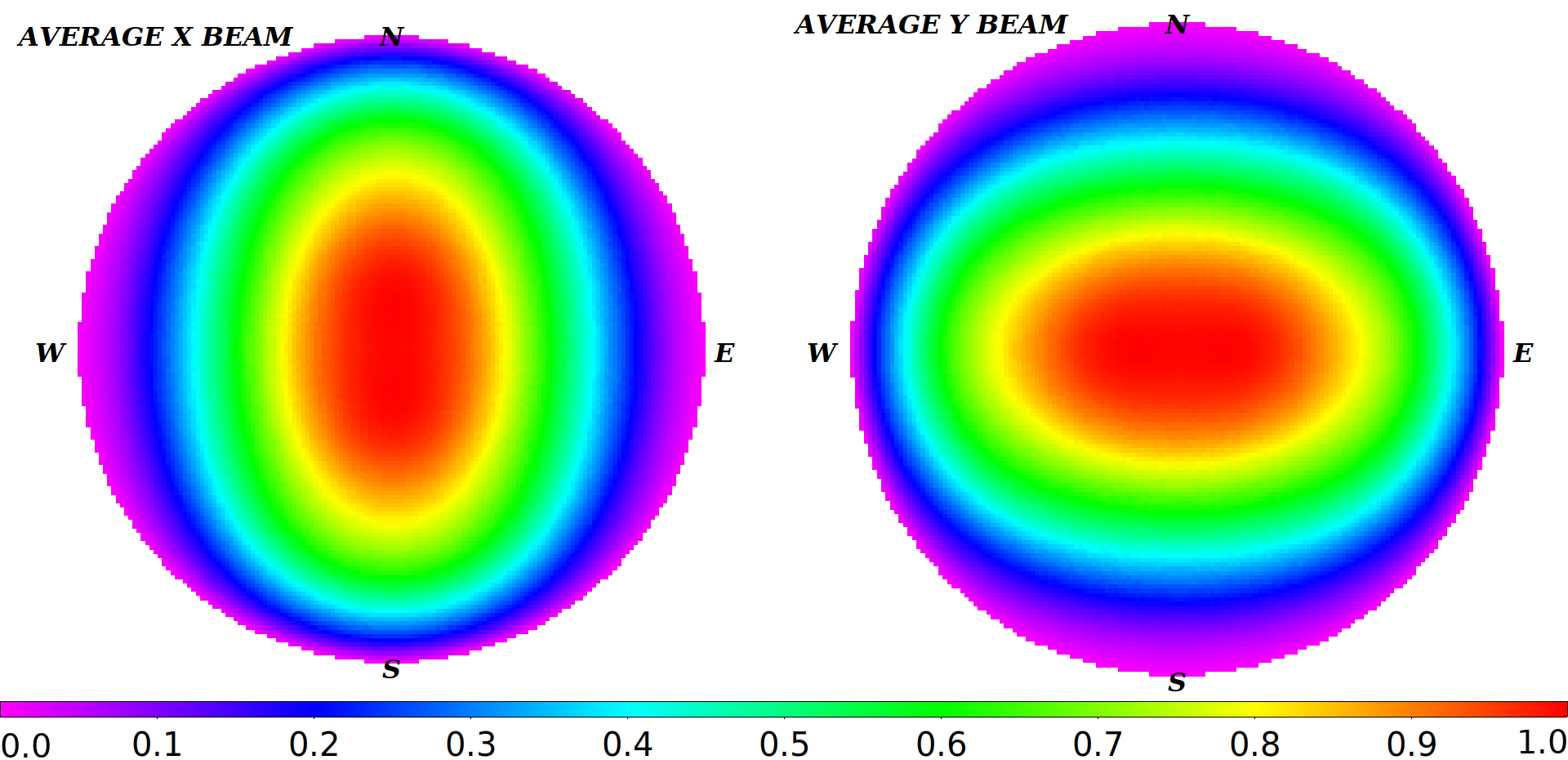}
	\caption{Examples of average beam patterns of the EDA-2 dipole in X polarisation (left) and Y polarisation (right) at 159.375\,MHz. These images were used to correct the original XX and YY images for the primary beam response if  correct flux scale was required to generate flux-calibrated lightcurves.}
    \label{fig_beam_xxyy}
\end{figure*}

\begin{figure}[h!]
    \includegraphics[width=\columnwidth]{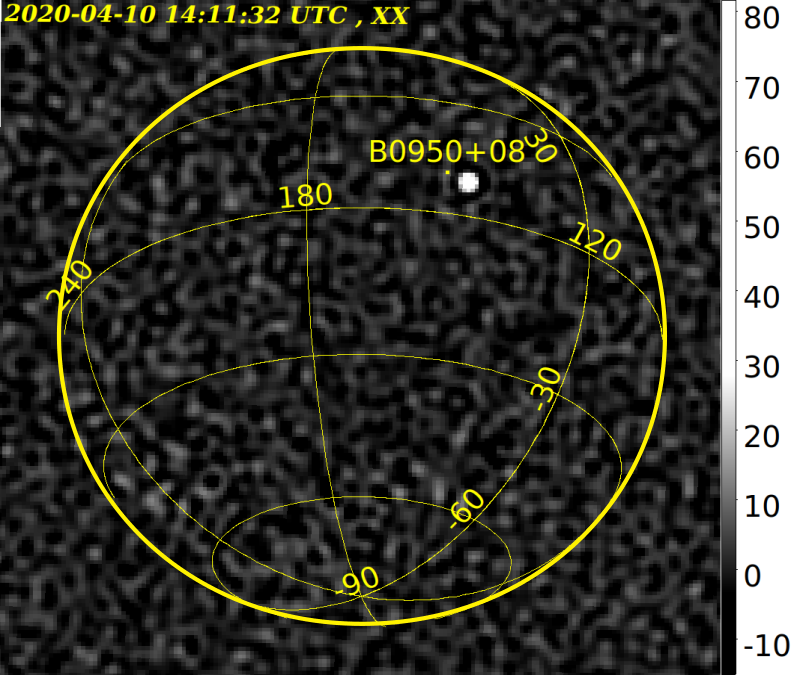}
	\caption{An example of 2\,s Stokes I difference image obtained by subtracting image started on 2020-04-10 at 14:11:30 UTC from the next image started at 14:11:32 UTC. The very bright ($\approx$80\,Jy) transient from pulsar PSR B0950+08 is clearly visible under the B0950+08 label. \red{The thicker yellow circle represents the horizon.}}
    \label{fig:eda2_allsky_diff_image}
\end{figure}

Visibilities XX and YY from each UV FITS file are imaged with \textsc{invert} task using robust=-0.5 weights \red{(no CLEAN was performed)}. \red{We note that all baselines were used and no $u,v$ cut was applied for imaging (only in calibration).} The all-sky image size is calculated as $N_{px} = O\pi D \nu / c$ pixels, where D is the station diameter (35\,m and 38\,m for the EDA2 and AAVS2 respectively), $\nu$ is observing frequency, c is the speed of light and the factor of $O$ comes from required over-sampling of the beam and is typically set to $O=3$ (e.g. 180$\times$180 pixels at 159.375\,MHz). The XX and YY images (examples in Fig.~\ref{fig_allsky_image_example}) can be divided by the corresponding images of the average embedded element beam (Fig.~\ref{fig_beam_xxyy}), but this step was only used when flux calibrated lightcurves were generated as artefacts introduced by the inaccuracies of the beam model can affect difference image-based transient searches.
In the next step XX and YY images are averaged to form Stokes I images\footnote{\red{Strictly speaking, when no beam-correction was applied they should be called pseudo-Stokes I images but we skipped this for brevity}}, which is the starting point for the presented ``blind'' transient searches. An example Stokes I all-sky image generated by the pipeline at $159.375$\,MHz is shown in right panel of Fig.~\ref{fig_allsky_image_example}. 
The all-sky Stokes I images are used in difference imaging procedure, where n-1 image is subtracted from the n$^{th}$ image and the resulting difference images (example in Figure~\ref{fig:eda2_allsky_diff_image}) are analysed in order to identify transient candidates, i.e. find pixels exceeding a predefined threshold of typically 5 standard deviations of the noise ($5\sigma_n$); details will be provided in Section~\ref{sec:realtime_transient_monitor}.

\red{The sensitivity in terms of System Equivalent Flux Density (SEFD) or effective area divided by system temperature (A/T) was measured from the noise in 0.14\,s difference images and compared against electromagnetic simulations and SKA-Low specifications \citep{EuCAP2021_paper}. These comparisons show that, especially at frequencies used in this paper 159.4, 229.7 and 320.1, the measured sensitivity values match the simulations very well at the time of the calibration (also single was applied to long observations) and differs by at most 20 -- 30\% a few hours apart from the time of the calibration. These discrepancies are mainly caused by gain variations related to changes in ambient temperature. Furthermore, we verified that the noise in difference images has Gaussian distribution. We also calculated standard deviations of these distributions using from the night 2020-04-10 (at 159.375\,MHz) and found them to be approximately 3.6\,Jy and 4.2\,Jy for the EDA2 and AAVS2 stations respectively. We performed sensitivity simulations as described by \citet{EuCAP2021_paper}, which predicted sensitivity averaged over 24\,h (changes with LST) to be around 2\,Jy in 2\,sec images and assuming 0.926\,MHz observing bandwidth for both stations. The discrepancy of about factor of 2 has not been fully understood. However, there are several differences with respect to the analysis presented in \citep{EuCAP2021_paper}, for example 2\,s dirty images vs. 0.14\,s lightly cleaned images, which can introduce some systematic effects and we leave further analysis to the future work.}

Even at the highest observing frequencies $\approx$320.3\,MHz with pixel size $\approx$0.365\degree, the sidereal sky movement of $\approx$0.00833\degree \, over the integration time of 2\,s corresponds to $\approx$2.2\% of the pixel size. This means that to produce a $\gtrsim$11\,Jy false candidate due to an artefact of the image subtraction originating from the sky movement (i.e. flux ``spill-over'' to a neighbouring pixel) a $\gtrsim$500\,Jy source is required. \red{The maximum ionospheric offsets reported at 170 -- 200\,MHz by \citet{2015GeoRL..42.3707L} were of the order of 1.2\,arcmin (below 1\,arcmin under normal conditions), which is a similar fraction of pixel ($\sim$2.8\%) and could cause similar effects if not the fact that the reported variability timescale was at the level of minutes (much longer than 2\,s images). Finally, candidates in difference images may also be caused by the source noise from very bright radio-sources, as recently reported by \citet{2021arXiv210111851M}. These effects justify the selection criteria excluding the regions around bright sources, such as the Sun, A-team sources, Galactic Plane and Bulge, from the algorithm (as the criteria $1-4$ in Sec.~\ref{subsec:filtering_transients}).}

\section{REAL-TIME TRANSIENT MONITOR}
\label{sec:realtime_transient_monitor}

\red{The all-sky images are produced in real-time and are immediately picked-up by the transients identification algorithm. We note, however, than at these early stages all the datasets were re-analysed off-line as the pipeline and algorithm has been undergoing very rapid development}. The algorithm for transient identification has a few stages, which will be described in this section. 

\subsection{SOURCE FINDING AND TRANSIENT DETECTION}
\label{subsec:source_finding}

For each difference image, the standard deviation $\sigma_{n}$ of the noise in the center of the image is calculated.
In the present version a simple source finding technique was implemented, but in the future it may be replaced by one of many existing source finding packages. 
All the pixels exceeding a specified threshold of 5$\sigma_{n}$ are identified and nearby pixels are grouped together by selection of only the brightest pixel within a 5-pixel radius in order to avoid multiple detections of the same source. 

\subsection{FILTERING TRANSIENT CANDIDATES}
\label{subsec:filtering_transients}

The 5$\sigma_{n}$ transient candidates identified in the difference images from each station are initially filtered by the following station-level criteria implemented to excise false candidates due to artefacts from imperfections of difference imaging around bright radio sources:

\begin{enumerate}
    \item \textbf{Galactic latitude} - candidates with Galactic latitude $|b| < 10$\,degrees are discarded.
    \item \textbf{Galactic Bulge} - candidates with $|b| < 15$\degree\, and Galactic longitude $|l| < 25$\degree\, are also discarded. Both ``Galactic coordinates'' cuts are similar to those used by \citet{2020PASA...37...39T}.
    \item \textbf{Angular distance to Sun} - candidates in angular distance from the Sun smaller than $8$\degree\, ($\sim 2-4$ beam sizes) are discarded. 
    \item \textbf{A-team sources} - candidates closer than $4$\degree \,from very bright A-team sources (Centaurus A, Hercules A, Hydra A, Pictor A, 3C444, Fornax A and Virgo A) are discarded.
\end{enumerate}

The station-level criteria were designed to excise only the most common and obvious sources of false alerts and save the list of transient candidates identified by each station to text files for further processing, filtering and post-processing analysis (including coincidence between the stations). However, some of these criteria may be relaxed in the future as the classification and filtering are improved. 

In the analysis described in this paper the source was required to be detected by both stations within a specified time window and angular distance - this requirement will be referred to as the coincidence. If the stations observed at the same frequency, the time window was set to the integration time (currently 2\,s). Otherwise, the time window was determined by the maximum dispersion measure (DM) of a potential transient, which was typically set to DM = 1000 $\text{pc/cm}^{3}$ corresponding to time window of $\approx 84.7$\,s when the stations observed at $159.375$ and $229.6875$\,MHz (or $\approx 122$\,s for observations at $159.375$ and $320.3125$\,MHz). The angular distance between transient positions in the images from both stations was required to be smaller than $3.3$\degree (corresponding to a station beam size at 150\,MHz). The candidates accepted by the coincidence requirement were saved to a log file. 
The candidates detected by both stations were further filtered by the following criteria in order to flag some other sources of false alerts:

\begin{enumerate}
\setcounter{enumi}{4} 
\item \textbf{Elevation cut} - candidates below certain elevation limit (default $25$\degree) are discarded in order to avoid RFI from ground-based FM, DTV and other transmitters in the population centers surrounding the MRO, such as Geraldton (South-West from the MRO), which is in a distance of $\approx$7 horizons away and can still be detected at the MRO at FM and DTV frequencies especially in favourable propagation conditions, i.e. tropospheric ducting \citep{2017rfi..confE...1S, 2020PASA...37...39T}. 
\vspace{0.1in}

\item \textbf{Catalogue of satellites} - each of the remaining candidates in the image is verified against the list of known satellites, which are above the horizon at the MRO at the time of the image (even up to a few hundred during a 2-second integration). In order to achieve this, every day Two-Line Element (TLE) catalogues are downloaded from the Internet sources (e.g. www.space-track.org) and a TLE file with all the satellites ($\sim$16000) is compiled for each day.
Then for each image the \textsc{sattest} software \citep{2008PhDT.......442S} generates a list of satellites with elevation $e > 0$\degree\, at the time when the image was collected.
Next, the position of each transient candidate is verified against this list and if a satellite closer than $4$\degree\, is found, the candidate is flagged with its NORAD ID from the TLE file. The satellites cross-matching radius $R_{sat}=4$\degree\, was selected based on a distribution of angular distances of transient candidates to the closest object from the TLE database, and as a compromise between efficiently flagging candidates due to TLE-objects and not excising all astrophysical transients because of false random cross-matches.

The number of TLE-satellites ($N_{sat}$) at elevations $\ge$25\degree \, at the time of each 2-second image is typically between $\approx$100 and 1000. Given that, the cross-matching radius $R_{sat}=$4\degree, and assuming an isotropic distribution of satellites\footnote{As can be seen from Figure~\ref{fig:sat_above_horizon_all} this assumption is only a very rough approximation}, the probability of randomly matching a transient candidate to one of these satellites (eq.~\ref{eq_sat_prob}) is around 50\% when only about 120 TLE-satellites are cross-matched. However it reaches $\approx$100\% for more than 240 TLE-satellites, and is below 1\% when the number of TLE-satellites at elevations $\ge$25\degree \, is $N_{sat}\lesssim$3 (cf. Sec.~\ref{subsec:results_satellites}).
Hence, the chances of random associations can be very high when all the TLE satellites are considered for cross-matching. This criteria is still under consideration whether only Low Earth Orbit (LEO) objects at a distance $\le$2000\,km should be considered due to the low chances of receiving reflected signals from further objects.
However, distant transmitting satellites could easily be detected. Therefore, so far we have been using the full list of satellites, not just LEO objects, but the criteria will be revised as more bandwidth and better time resolution will improve chances for automatic classification of moving objects.
\vspace{0.1in}

\item \textbf{Catalogue of bright radio sources} - each of the candidates (including those flagged as a TLE satellite) is verified against a catalogue of bright radio-sources (larger than the short list of A-team sources). If the candidate is closer than 4\degree\, from the source in the catalog it is flagged with the name of this sources and the angular distance to it is also saved to the log file.
\vspace{0.1in}

\item \textbf{Excision of images with a bright RFI transient} - if a very bright RFI transient (flux density $\ge$ 300 Jy/beam) is identified in the images from both stations, then all the candidates from these images are rejected. This is to reject false candidates from very strong side-lobes resulting from such a bright RFI event, which can cause many false candidates across an entire image.
\vspace{0.1in}

\item \textbf{Sun / daytime} - the images collected when the Sun was above elevation of 20\degree \, are excluded from the analysis due to very strong side-lobes from the Sun (compare to the previous criterion). 
\vspace{0.1in}

\item \textbf{Pre-defined flight paths} - candidates which are less than 10\degree \, from two typically used flight paths, which were fitted to moving candidates from one of the analysed nights (details in Sec.~\ref{subsec:results_aircraft}), are excised. This criterion can be extended in the future to use the actual data from the plane tracking websites in order to unambiguously excise false candidates due to RFI from planes. 
\vspace{0.1in}

\end{enumerate}

The candidates not rejected by the above criteria are saved to the transient candidates log file for further inspection, whilst the rejected events are saved to a separate log file.

\section{PRELIMINARY RESULTS}
\label{sec:results}

After applying all the criteria described in Section~\ref{subsec:source_finding} the transient candidates matched to TLE-catalogue satellites or A-team radio-sources are flagged and the corresponding NOARD ID or/and radio-source name and the angular distance are saved to a log file. Due to large number of known satellites (up to 1000 when any distance is considered) above the horizon, a large fraction of the TLE catalogue satellites cross-matches are false, which unfortunately noticeably reduces efficiency of the algorithm to detect astrophysical transients.
Table~\ref{tab_candidates} shows the number of candidates after subsequent criteria for \Ncoincnights analysed nights ($\approx$\Ncoinctime hours in total) when data from both stations were collected simultaneously. The final list of candidates not matched to any satellite required further visual inspection. As can be seen from Table~\ref{tab_candidates} in some cases it was impossible to visually inspect all of them.
However, at this stage we are not intending to do it as the number of candidates will be significantly reduced when fine channelised images, larger frequency band and better time resolution are implemented, which will also help in excision of moving objects.

\red{Using 160\,MHz images (180$\times$180 pixels), we verified that the number of pixels above the minimum elevation of 25\degree\, is approximately 14261. Given the probability of exceeding the $5\sigma_n$ threshold by the Gaussian noise is $\approx 2.86 \times 10^{-7}$, and the requirement for the candidate to be detected by both stations at the same sky position, the expected number of false candidates in 43200 images from 24\,h of observations is $\approx$0.7. Moreover, this very low number is before any criteria other than the coincidence and elevation cut. We confirm, that we inspected all the events in the column 9 of Table~\ref{tab_candidates}, and none of them looked like caused by fluctuation of the noise with most of them having SNR$\gtrsim10$.}

\begin{table*}
\caption{Number of candidates after the main filtering criteria for \Ncoincnights analysed nights when both stations were collecting data. For some columns two numbers are shown for the EDA2 and AAVS2 respectively. The columns $\text{N}_\text{eda2}$ and $\text{N}_\text{aavs2}$ are numbers of transients detected in difference images from the EDA2 and AAVS2 stations respectively. $\text{N}_\text{coinc}$ is the number of candidates after requiring time coincidence, i.e. maximum of 2\,s or dispersion time corresponding to maximum DM=1000\,pc/cm$^3$ if the stations observed at different frequencies, and spatial coincidence in radius $R_{coinc} = 3.3$\degree \, between both stations. $\text{N}_\text{sat}$ is the number of candidates matched to satellites in TLE catalogue according to the criteria described in the text. $\text{N}_\text{acc}$ is the number of the remaining candidates (excluding the transients associated with PSR B0950+08 shown in the separate column $\text{N}_{\text{B0950}}$), which are potentially of astrophysical origin.}

\centering
\begin{tabular}{@{}ccccccccccc@{}}
\hline\hline
Start Date & Frequencies$^a$    & Observing & $\text{N}_{\text{eda2}}$ & $\text{N}_{\text{aavs2}}$ & $\text{N}_{\text{coinc}}$ & $\text{N}_{\text{sat}}$ & $\text{N}_{\text{planes}} $ & $\text{N}_{\text{cand}}$$^{b}$ & $\text{N}_{\text{acc}}$ & $\text{N}_{\text{B0950}}$  \\
  (UTC)    &  (MHz) &      interval      &            &             &             &           &        &             &            & \\
           &        &       (hours)      &            &             &             &           &        &             &            & \\
\hline%
 2020-04-10  & 159.4 / 159.4 & 22.80 & 34728   & 37097  & 1624 & 549 & 210 & 2 & 0 & 180 \\ 
 2020-04-11  & 229.7 / 229.7 & 33.16 &  59587  & 105599 & 777 & 244 & 111 & 38 & 4$^c$ & 0  \\ 
 2020-04-16  & 320.3 / 320.3 & 23.00  &  29172 & 188386 & 12$^c$ & 0   & 0   & 0 & 0       & 0 \\
 2020-04-29  & 159.4 / 159.4 & 10.21  & 16132  & 22545  & 1650   & 336 & 201 & 6 & 1$^{d}$ & 0 \\ 
 2020-05-30  & 159.4 / 229.7 & 10.9 & 46155 & 43564 & 1108 & 25 & 0 & 26 & 22$^e$ & 0 \\ 
 2020-06-26  & 159.4 / 159.4 & 131.47 & 38552 & 22163 & 1682 & 797 & 113 & 9 & 0 & 27 \\ 
 2020-07-07  & 159.4 / 229.7 & 23.91 & 24799 & 64990 & 270 & 75 & 9 & 6 & 0 & 0 \\ 
 2020-07-09  & 159.4 / 229.7 & 1.00 & 483 & 240 & 11 & 9 & 0 & 0 & 0 & 0 \\ 
 2020-09-11$^f$  & 159.4 / 229.7 & 2.84 & 582 & 83428 & 0 & 0 & 0 & 0 & 0 & 0 \\ 
 2020-09-14  & 159.4 / 229.7 & 32.25 & 78630 & 67388 & 28 & 27 & 20 & 1$^g$ & 0 & 0 \\ 
 2020-09-18  & 159.4 / 229.7 & 25.69 & 36425 & 51836 & 31 & 31 & 0 & 0 & 0 & 0 \\
 2020-09-25  & 159.4 / 229.7 & 32.66 & 43471 & 138642 & 83 & 76 & 16 & 0 & 0 & 0 \\
 2020-09-27  & 159.4 / 312.5 & 15.59 & 16482 & 61083 & 0 & 0 & 0 & 0 & 0 & 0 \\
 2020-10-01  & 159.4 / 312.5 & 0.78 & 313 & 54427 & 0 & 0 & 0 & 0 & 0 & 0 \\
\hline\hline
\end{tabular}
\newline
\begin{flushleft}
\tabnote{$^a$ Frequencies are approximated to a first decimal digit with the exact frequencies 159.375, 229.6875, 312.5 and 320.3125\,MHz \red{at these frequencies the synthesised beam sizes are approximately 2.4\degree (2.2\degree), 1.7\degree (1.5\degree), 1.1\degree\, and\,1.2\degree(1.1\degree)\,respectively, where values in brackets are for the AAVS2 station (except 312.5\,MHz observed only with AAVS2 station).}} 
\tabnote{$^b$ The number of astrophysical candidates after excluding candidates caused by the PSR B0950+08 pulses}
\tabnote{$^c$ These candidates were discarded upon visual inspection as satellites, planes or artefacts }
\tabnote{$^d$ Number of candidates which passed visual inspection, but could not be confirmed to be of astrophysical origin}
\tabnote{$^e$ Multiple transients from the same position, which is currently under investigation and will be reported in a future publication}
\tabnote{$^f$ The EDA2 was only collecting data for about 3\,hours, but more AAVS2 data were used for PSR B0950+08 monitoring}
\tabnote{$^g$ Rejected as a moving object (most likely a satellite) after visual inspection of images }
\end{flushleft}
\label{tab_candidates}
\end{table*}

\subsection{RADIO FREQUENCY-INTERFERENCE}
\label{subsec:results_rfi}

\subsubsection{SATELLITES}
\label{subsec:results_satellites}

\begin{figure*}[h!]
    \includegraphics[width=\textwidth]{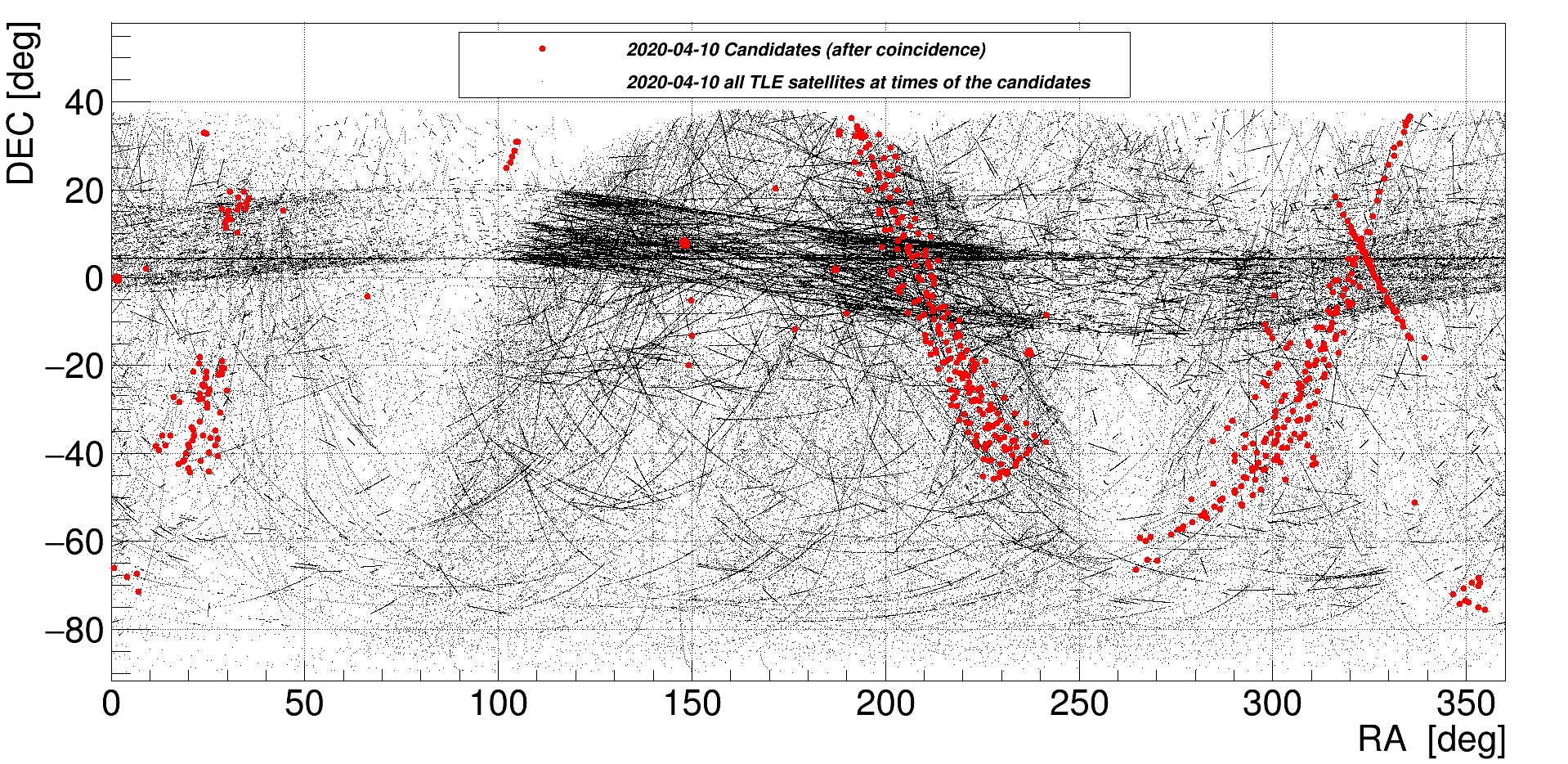}
	\caption{Distribution of all candidates detected in the 2020-04-10/11 data (red dots) and positions of all satellites above the horizon at the MRO for all the corresponding timestamps (small black dots). The black dots form clear patterns, such as for example geo-stationary satellites in approximately 20\degree \, wide belt of objects around the Equator. The observed transients detected from PSR B0950+08 form a grouping of red dots at ($\lambda$,$\delta$) $\approx$ (150\degree,10\degree) and this is how these transients were first noticed amongst the other transient candidates.}
    \label{fig:sat_above_horizon_all}
\end{figure*}


The positions of identified transient candidates were compared to positions calculated for objects in the TLE database (typically about 16600 objects) using the \textsc{SATTEST} program. Figure~\ref{fig:sat_above_horizon_all} shows distribution of all transient candidates identified in the 2020-04-10/11 data over-plotted with calculated positions of all satellites above the horizon at the MRO at the times of the identified transients. The patterns in expected orbital positions of TLE-satellites are clearly visibile (e.g. geo-stationary satellites forming an approximately 20\degree \, wide strip of objects around the Equator). It was also verified that over the 24\,hour interval starting at around 21:30 AWST on 2020-04-10 the number of satellites at elevations $\ge$25\degree\, in an arbitrary distance from the Earth was between 860 and 1010, while number of only LEO satellites (height $\le$2000\,km) was between 85 and 170. Given the number of TLE-catalogue satellites $\text{N}_{sat}$ above elevation $\text{e}_{min}$ at a particular moment, the excision radius $\text{R}_{coinc}$ and minimum considered elevation $\text{e}_{min}$ the probability, p, of randomly matching a TLE-catalogue satellite to a transient candidate can be calculated as:

\begin{equation}
p = N_{sat} \frac{sin^2(\frac{R_{coinc}}{2})}{sin^2(\frac{90 - e_{min}}{2})}.
\label{eq_sat_prob}
\end{equation}

Assuming uniform distribution of satellites, which as Fig.~\ref{fig:sat_above_horizon_all} shows is not an ideal approximation, and  equation~\ref{eq_sat_prob} results in the probability (p) of falsely matching a transient candidate to a TLE satellite greater than one (between 3.6 and 4.3) for satellites at an arbitrary distance from the Earth, and 0.36 - 0.72 for LEO satellites (mean 0.54).
These probabilities are in an order-of-magnitude agreement with our analysis. When the transient candidates were cross-matched against TLE satellites in an arbitrary distance from the Earth, then $\approx$92\% of transient candidates were matched to a TLE-satellite. The disagreement ($>$100\% probability expected vs. 92\% probability observed) is most likely due to the fact that the satellite positions are not isotropically distributed and are clustered around certain orbits (e.g. geo-stationary) and because of this clustering the probability of matching transients at any sky coordinates is lower than predicted for an isotropic distribution of satellites. Therefore, many of the the real astrophysical transients from PSR B0950+08 were not excised by this criterion. On the other hand, when only LEO satellites were cross-matched, the percentage of transient candidates matched to LEO satellites was approximately 55\%, which is close to the predicted value of 54\%, indicating that for these objects the isotropic distribution assumption is indeed valid.
The full description of these probabilities requires more simulation work and is beyond the scope of this paper, but will be performed if required by the future analysis.
Clearly, excision of satellites using a catalogue of orbital elements is not an optimal approach, but it may not be required once the increased bandwidth and better frequency and time resolutions become available.

\begin{figure}[h!]
    \includegraphics[width=\columnwidth]{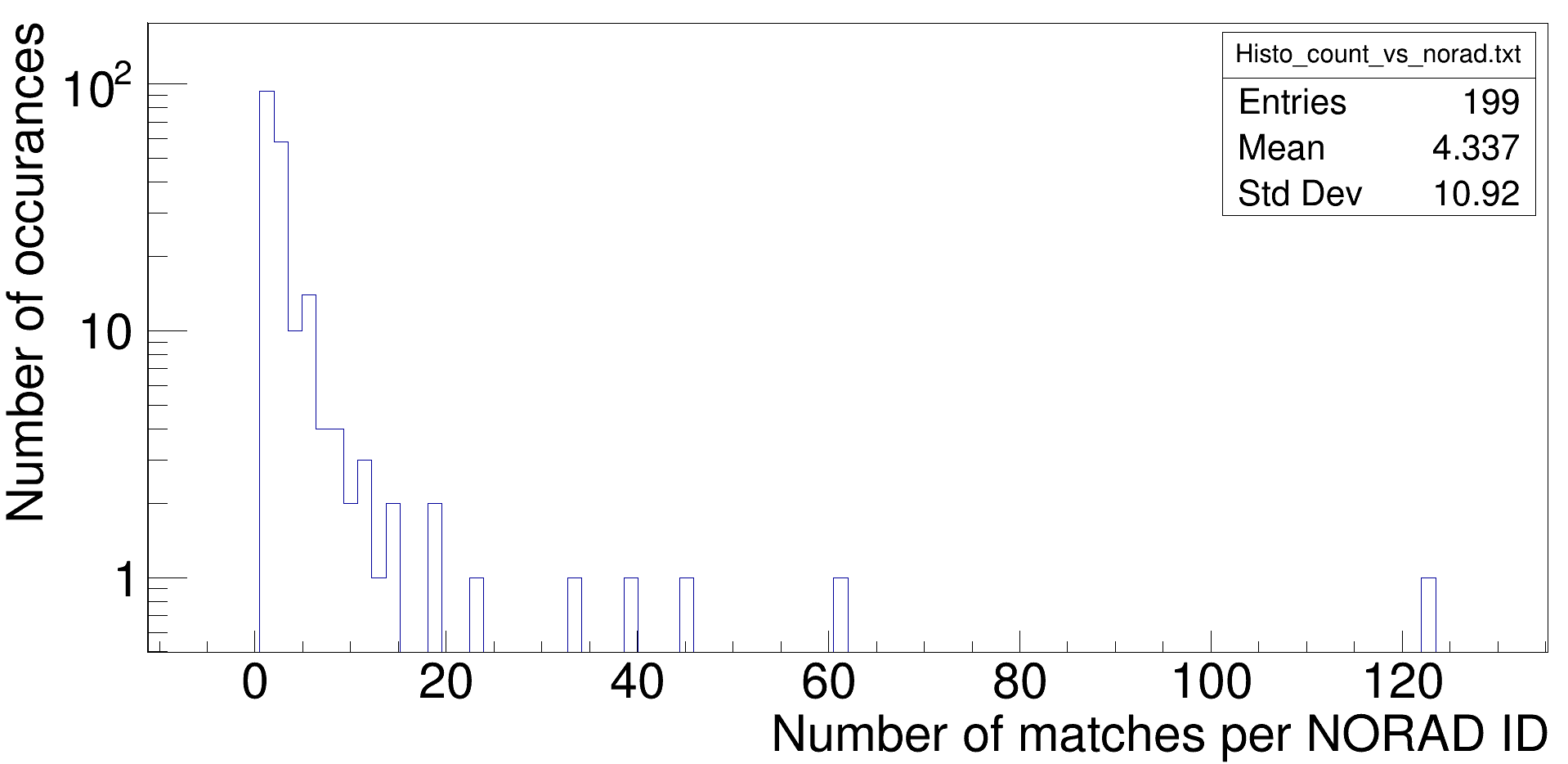}
	\caption{Number of transient candidates matches per NORAD ID for the data from night 2020-04-10/11. The peak at 123 corresponds to BGUSAT (NORAD ID 41999).}
    \label{fig_matches_per_noradid}
\end{figure}

\begin{figure}[h!]
    \includegraphics[width=\columnwidth]{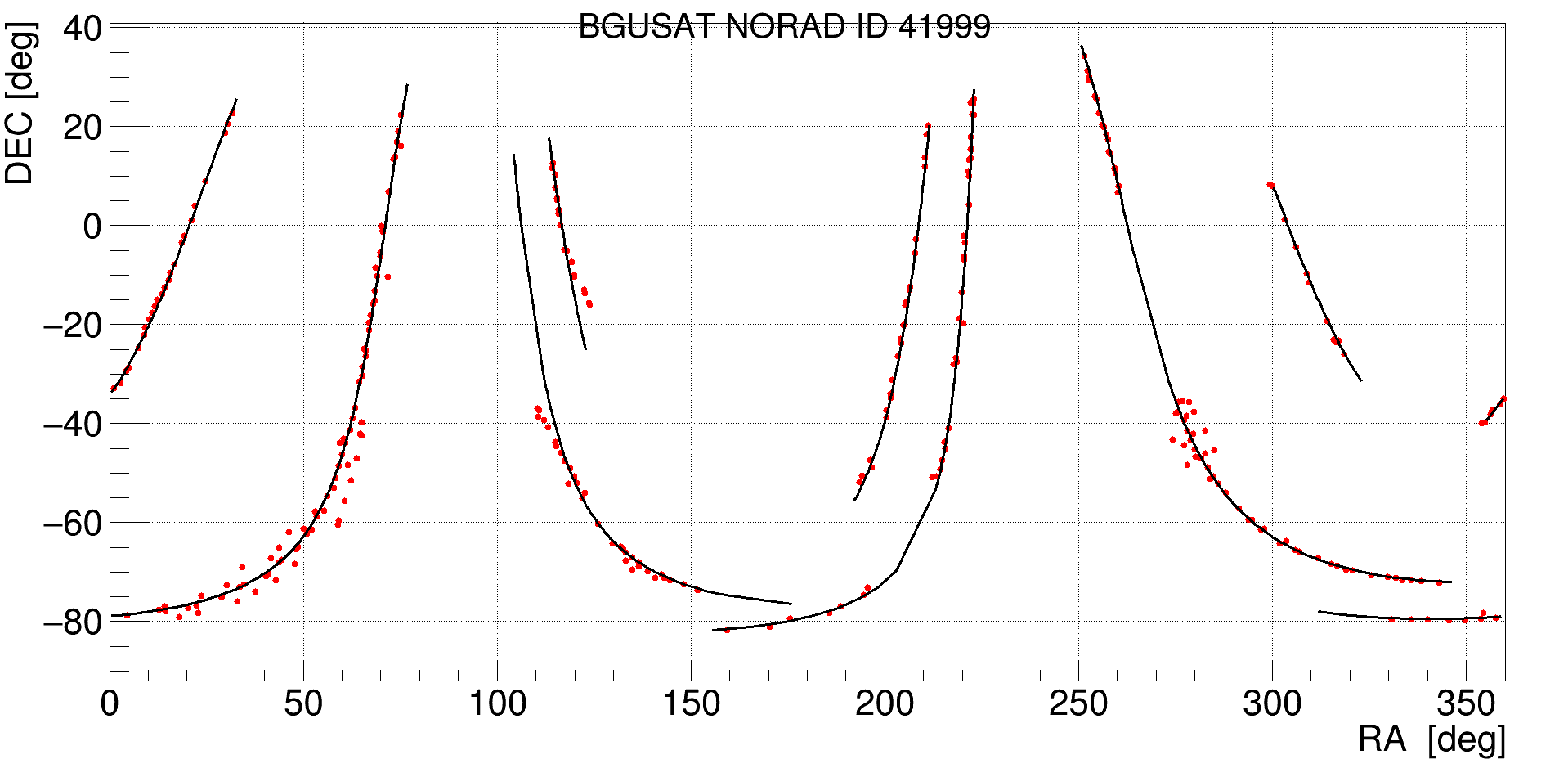}
	\caption{Example detections of BGUSAT (NORAD ID 41999) in AAVS2 difference images at 159.375\,MHz (red points) and predicted paths (black curves). Several few minute-long passages were observed between 2020-06-26 21:15:42 and 2020-07-02 08:44:05 AWST. In order to create this image the transient searching algorithm was executed without any restrictions on the Sun elevation.}
    \label{fig:bgusat_passages}
\end{figure}

Finally, it can be calculated (eq.~\ref{eq_sat_prob}) that the probability of matching a specific satellite (with a given NORAD ID) to a transient candidate is very low $\sim$0.3\% and therefore the probability of randomly matching the same satellite more than 5 times is $\lesssim10^{-13}$.
Hence, it can be assumed that multiple matches to the same satellite are genuine identifications of this satellite. Fig~\ref{fig_matches_per_noradid} shows the number of satellite observations per unique NORAD ID for the data from night 2020-04-10/11. The distribution peaks at a very low number of matches and falls-off rapidly indicating that the majority of cross-matches are random false-identifications. However, the peak at 123 matches is due to BGUSAT (NORAD ID 41999), which has been confirmed to be a genuine detection.
 BGUSAT is a nanosatellite which has been regularly detected in the EDA2 and AAVS2 data at many observing frequencies and it is most likely transmitting out of its nominal band \citep{2020PASA...37...39T}.
 Example detections of BGUSAT at 159.375\,MHz are shown in Figure~\ref{fig:bgusat_passages}. Other NORAD IDs with more than 10 matches are summarised in Table~\ref{tab_norad_matches}. This table shows that, while there are satellites (like BGUSAT) \red{likely} transmitting at wide range of frequencies 159.4 and 229.7 MHz and in FM band \citep{2020PASA...37...39T}, in general a completely different group of satellites have been detected at frequencies above 229.7\,MHz (mostly Russian COSMOS class satellites). We also note that no satellites were identified when both stations observed at about 320.3\,MHz (2020-04-16/17 data), which seems to be a very clean band to look for transients. The objects in Table~\ref{tab_norad_matches} are different from those detected with the MWA in $72.335 - 103.015$\,MHz band (partially overlapping with the FM band) by the earlier studies \citep{2020PASA...37...52P,2020PASA...37...10P,2018MNRAS.477.5167Z}.

\subsubsection{ORIGIN OF THE SIGNALS}
\label{subsec_sat_signals_origin}

\red{Given large uncertainties in Radar Cross-sections (RCS) for the majority of the detected objects, the full analysis of the detected signals being due reflections or transmissions is beyond the scope of this paper. A possible source of signals for these reflections at the analysed frequencies are ground-based transmitters in Western Australia or possibly beyond (subject to power constraints). Nevertheless, out of all the frequency channels used in the presented analysis only the frequency channel 229.6875\,MHz is within the frequency band of DTV transmitters in Australia, which extends up to 230\,MHz (see for example Figure 2 in \citet{2016arXiv161004696S}). Specifically there are 50\,kW DTV transmitters in Perth covering frequency band 170 -- 230\,MHz. None of the other frequency channels used in this analysis are within the frequency bands allocated for broadcasting (see Australian Communications and Median Authority (ACMA)\footnote{https://www.acma.gov.au/australian-radiofrequency-spectrum-plan}). Hence, the potential reflections could not be due to DTV or FM transmitters. The signal sources for the potential reflections at frequencies other than 229.6875\,MHz could be located outside Australia, but we did not explore the list of frequencies and transmitters in the nearest countries. Nevertheless, we provide a simple estimate of the expected flux densities. Assuming isotropic gain of ground-based DTV transmitters of power $\text{P}_{\text{tr}}^{\text{kW}}$ (in kW) radiating uniformly over the $\text{BW}_{\text{MHz}}^{\text{tr}}$ band (7\,MHz for DTV) as source of the reflected signal, equal distance $\text{r}_{\text{km}}$ (in km) from transmitter to receiver in a bi-static radar configuration, and using a textbook radar equation, the following equation can be derived to estimate expected flux density due to reflections:}
\begin{equation}
\text{f}_{\text{r}} \propto \text{4.5\,[mJy]} \left( \frac{\text{1000}}{\text{r}_{\text{km}}} \right)^4 \left( \frac{\text{P}_{\text{tr}}^{\text{kW}}}{\text{50 kW}} \right) \left( \frac{\text{7\,MHz}}{\text{BW}_{\text{MHz}}^{\text{tr}}} \right) \text{RCS},
\label{eq_flux_reflected}
\end{equation}
\red{where RCS is in m$^2$. Given that flux densities observed for the objects in Table~\ref{tab_norad_matches} are in the range from tens to a few thousands Jy, it is nearly impossible that they are caused by off-the-satellites reflections of signals emitted by DTV or other ground-based transmitters in Western Australia or further. Using BGUSAT at the height about 500\,km as an example, the expected flux density from a 50\,kW transmitter in Perth would be below 7.2\,mJy, whilst the observed flux densities of BGUSAT were in the range 10 -- 3500\,Jy. The observed flux densities are much more consistent with line-of-sight propagation from a low power transmitter ($\le$1\,W) with a small fraction of out-of-band ``spill-over'' over a wide frequency band.}

\begin{table*}
\caption{A list of TLE objects most commonly detected with the system and having at least 10 matches (except a few special cases). In order to create this table, the filtering criteria were relaxed to allow daytime and minimum elevation of transient candidate of 15\degree \,(nominally transients are searched in nighttime data only and at elevations $\ge$25\degree). Additional information in columns 5, 6 and 7 were obtained from \citet{2020PASA...37...39T}, the web pages \url{https://www.n2yo.com/satellite/?s=NORADID} and \url{http://www.zarya.info/Frequencies/FrequenciesAll.php}, where NORADID has to be replaced by a value from the first column of this table. For the rocket bodies, space debris and other inactive elements the 5-th column contains N/A value. \red{The values of Radar Cross-Section (RCS) in column 7 are poorly known and represent the best estimates we could find. At these frequencies the most likely sources of reflected signal are ground-based transmitters. However, except the frequency 229.6875\,MHz, they cannot be DTV or FM transmitters located in Australia (see discussion in Section~\ref{subsec_sat_signals_origin}).}}
\centering
\footnotesize
\begin{tabular}{@{}cccccccc@{}}
\hline\hline
\textbf{NORAD}  & \textbf{Satellite} & \textbf{Start} & \textbf{Number} & \textbf{Downlink}    & \textbf{Mission} & \textbf{RCS}  & \textbf{Approx.}   \\
\textbf{ID \#}   &   \textbf{Name}    &  \textbf{Date} & \textbf{of}   & \textbf{Frequencies} & \textbf{Status}   &  \textbf{[}$\textbf{m}^\textbf{2}$\textbf{]}  & \textbf{Height}      \\
                &                    & \textbf{[UTC]} &         \textbf{matches}  & \textbf{[ MHz ]}     &          &      &    \textbf{[km]}       \\
\hline%
\hline
41999 & BGUSAT              & 2020-04-10 & 148 & unknown & active & $<0.1$ & 500 \\
      &                     & 2020-04-29 & 150 &         &   &     \\ 
      &                     & 2020-06-26 & 499 &         &   &     \\ 
      &                     & 2020-07-07 & 15$^a$  &         &   &     \\ %
      &                     & 2020-09-14 & 13$^a$  &         &   &     \\ %
      &                     & 2020-09-18 & 14$^a$  &         &   &     \\ %
      &                     & 2020-09-25 & 45$^a$  &         &   &     \\ %
\hline
39427 & TRITON 1            & 2020-04-10 & 62  & 145.818 , 145.823 & inactive & unknown & 650 \\
      &                     & 2020-04-29 & 27  &         &   &     \\      
      &                     & 2020-06-26 & 191 &  & & \\
      &                     & 2020-09-25 & 3$^b$ &  & & \\
\hline
40547 & IRNSS-1D            & 2020-04-10 & 61 & unknown  & active & unknown & 36000 \\
      &                     & 2020-04-29 & 24 &          &   &     \\    
\hline      
40209 & ATLAS 5 CENTAUR R/B &  2020-04-10 & 39 & N/A & rocket body & unknown & 14600 -- 32500\\ 
33057 & ARIANE 5 R/B & 2020-04-10 & 21 & N/A & rocket body & 21.8 & 19700 -- 30000 \\ 
117   & SOLRAD 3/INJUN 1    &  2020-04-10 & 23 & unknown & inactive & 0.44 & 960 \\ 
28393 & AMAZONAS & 2020-04-10 & 18 & unknown & inactive & 23.6 & 36300 \\ 
26470 & NILESAT 102 & 2020-04-10 & 15 & unknown & retired & 27.9 & 36400 \\
26107 & ASIASTAR & 2020-04-10 & 15 & unknown & inactive & 20.0 & 35700 \\ 
41272 & NOAA 16 DEB & 2020-04-10 & 12 & unknown & debris & unknown & 900 \\
39988 & BREEZE-M DEB & 2020-04-10 & 14 & unknown & satellite debris & 0.003 & 12400 -- 16700 \\
38592 & BREEZE-M DEB & 2020-04-10 & 11 & unknown & satellite debris & 0.003 & 10800 \\
\hline
27868 & COSMOS 2400 & 2020-04-11 & 27 & 244.512 , 261.035 & unknown & 0.99 & 1500 \\
27059 & GONETS D1 8 & 2020-04-11 & 27 & 244.512 , 261.035 & unknown & 1.00 & 1400  \\
32955 & COSMOS 2438 & 2020-04-11 & 26 & 244.512 , 261.035 & unknown & 0.97 & 1500 \\
37153 & STRELA 3 & 2020-04-11 & 25 & 244.512 , 261.035 & unknown & 0.91 & 1500 \\
28420 & COSMOS 2409 & 2020-04-11 & 21 & 244.512 , 261.035 & unknown & 1.00 & 1500 \\
38733 & COSMOS 2481 & 2020-04-11 & 20 & 244.512 , 261.035 & unknown & 0.84 & 1500 \\
27465 & COSMOS 2391 & 2020-04-11 & 20 & 244.512 , 261.035 & unknown & 0.94 & 1500 \\
35500 & COSMOS 2453 & 2020-04-11 & 19 & 244.512 , 261.035 & unknown & 0.88 & 1500 \\
27056 & COSMOS 2385 & 2020-04-11 & 19 & 244.512 , 261.035 & unknown & 0.89 & 1400 \\
32956 & COSMOS 2439 & 2020-04-11 & 16 & unknown & unknown & 0.94 & 1500  \\ 
\hline
42778 & MAX VALIER  & 2020-06-26 & 17 & 145.860 & active & 0.1 -- 1 & 500 \\ 
7143 & DELTA 1 DEB  & 2020-06-26 & 14 & unknown & satellite debris & 0.07 & 1500 \\ 
22533 & THORAD AGENA D DEB  & 2020-06-26 & 13 & unknown & satellite debris & 0.013 & 1000 \\ 
16849 & COSMOS 1761  & 2020-06-26 & 11 & unknown & inactive & 12.14 & 22800 -- 32500 \\ 
29793 & FENGYUN 1C DEB  & 2020-06-26 & 10 & unknown & satellite debris & 0.018 & 1000 \\ 
11145 & OPS 9442 (DSCS 2-12) & 2020-06-26 & 9 & unknown & inactive & 5.04 & 36300 \\ 
18802 & COSMOS 1823 DEB  & 2020-06-26 & 9 & unknown & satellite debris & 0.06 & 1630 \\
\hline 
\hline
\end{tabular}
\newline
\begin{flushleft}
\tabnote{$^a$ It was observed that BGUSAT was much brighter ($\sim$30 -- 250 times) at 159.375\,MHz ($\sim$900\,Jy/beam) than at 230\,MHz ($\sim$30\,Jy/beam).}
\tabnote{$^b$ Only 3 detections, but shown here to exemplify another detection much brighter (nearly 230 times) at 159.375\,MHz ($\sim$34\,Jy/beam) than at 230\,MHz ($\sim$7700\,Jy/beam). }
\end{flushleft}
\label{tab_norad_matches}
\end{table*}

\subsubsection{AIRCRAFT}
\label{subsec:results_aircraft}

\begin{figure}[h!]
    \includegraphics[width=\columnwidth]{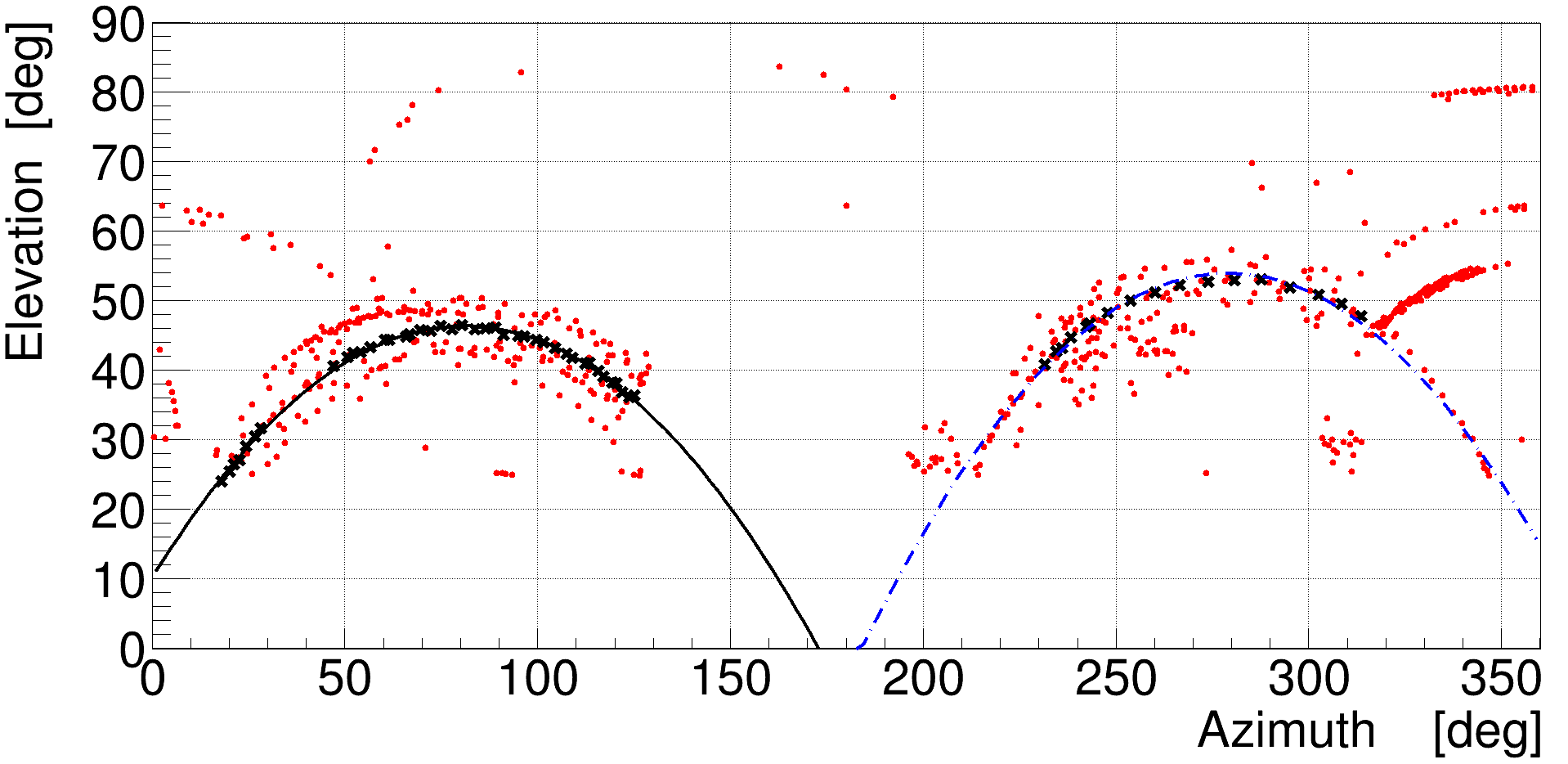}
	\caption{Transient candidates from 2020-04-10/11 data over-plotted with second order polynomials fitted to two paths. These parabolas were later used in excising RFI from aircraft moving along these routes.}
    \label{fig_parabolas}
\end{figure}

Bright transient candidates were identified to be mostly due to airplanes passing in the close vicinity of the MRO. Two main routes were identified and a parabola in elevation vs. azimuth was fitted to a set candidates from 2020-04-10/11 brighter than 300\,Jy/beam (black crosses in Fig.~\ref{fig_parabolas}). The relatively low threshold of 300\,Jy/beam was chosen to fit the curves to a sufficiently large number of points. These two parabolas were later used to excise transient candidates if they were closer than three times beam size ($\approx$10\degree). These is not an ideal criterion because the routes of planes may vary between days. Therefore, in the future we will try to automatically fit tracks to the moving objects, use time and frequency resolution (once coarse channel data are channelised and images in fine channels are formed), and if still required, use aircraft tracking services to obtain coordinates of aircraft in the vicinity of the MRO. \red{These signals may be caused by reflections off nearby (within tens of km) airplanes (with RCS$\gtrsim$2m$^2$). It can be shown, using equation~\ref{eq_flux_reflected}, that reflections of signals emitted over a narrow band ($\sim$10\,kHz) by even low power ($\le$1\,W) ground-based transmitters can cause high flux density candidates (of the order of thousands of Jy).}

\red{The distance between the two stations is approximately 165\,m. Therefore, the very near-Earth objects (like planes), can be observed in the images from two stations at slightly different positions with respect to stars due to parallax effect, which could be used to excise these kind of objects. We assume that the minimum measurable angular distance is $\sim$\sfrac{1}{5} of the synthesized beam ($\approx$2.3\degree\, and $\approx$1.15\degree\,at 160 and 320\,MHz respectively). Therefore, the maximum altitude to which parallax can be observed is 21 and 41\,km at 160 and 320\,MHz respectively (assuming a plane flying overhead). Since, our spatial coincidence radius (3.3\degree) is larger than the parallax angles at these frequencies we did not take advantage of this effect in the presented analysis. However, in the future if the positions of objects are indeed determined with the accuracy of at least $\sim$\sfrac{1}{5} of the synthesized beam it will be possible to use the parallax to excise objects closer than $\sim$41\,km. However, the efficacy of this criterion may be limited to aircraft flying over-head as for the objects closer to the horizon (distance to horizon is $\approx$700\,km for objects at altitude 40\,km) the parallax angles will be smaller than the achievable angular resolution.}

\subsection{PSR B0950+08}
\label{subsec:b0950}

The candidate events identified by the ``blind'' search algorithm described in Section~\ref{subsec:source_finding} and passing all the criteria  were visually examined, and a summary is given in Table~\ref{tab_candidates}. A majority of these candidates were seen as single image transients, which we could not assign to any of the satellites in the database, nor match to any pre-defined flight paths  (Section~\ref{subsec:results_aircraft}). A large grouping of these candidates near the position of PSR B0950+08 was identified in the 2020-04-10/11 data (Fig.~\ref{fig:sat_above_horizon_all} and Tab.~\ref{tab_candidates}), which we interpret as bright pulses from the pulsar and discuss in this section. We however note that our 2\,s integration time means averaging over approximately 8 rotation periods (P$\approx$0.253\,s), and hence these events are not individual bright pulses, although they necessarily imply that there were multiple bright individual pulses in the corresponding integration. We hence refer to these ``events'' as bright pulses. This also means that some of the individual pulses in a given integration are likely much brighter than that appear to be in our analysis.

\begin{figure*}[h!]
    \includegraphics[width=\textwidth]{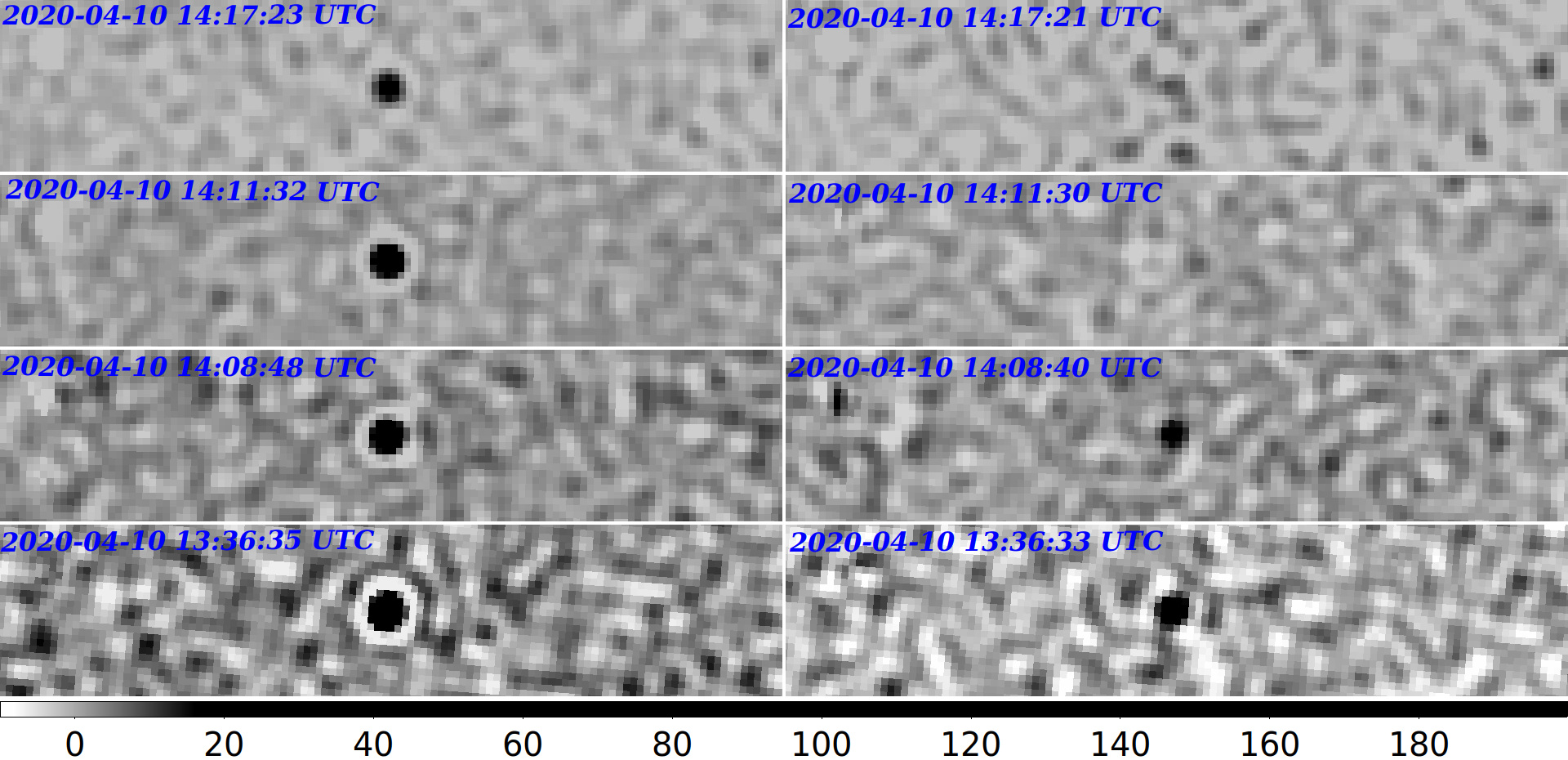}
	\caption{Stokes I difference images of the four brightest pulses from PSR B0950+08 2\,s (left column) and the preceding 2\,s images (right column). The images were obtained from Stokes I images after subtracting a running median of 30 images from the original images because it was very difficult to see the transients in the original images.}
    \label{fig_b0950_examples}
\end{figure*}

\begin{figure*}[h!]
    \includegraphics[width=\columnwidth]{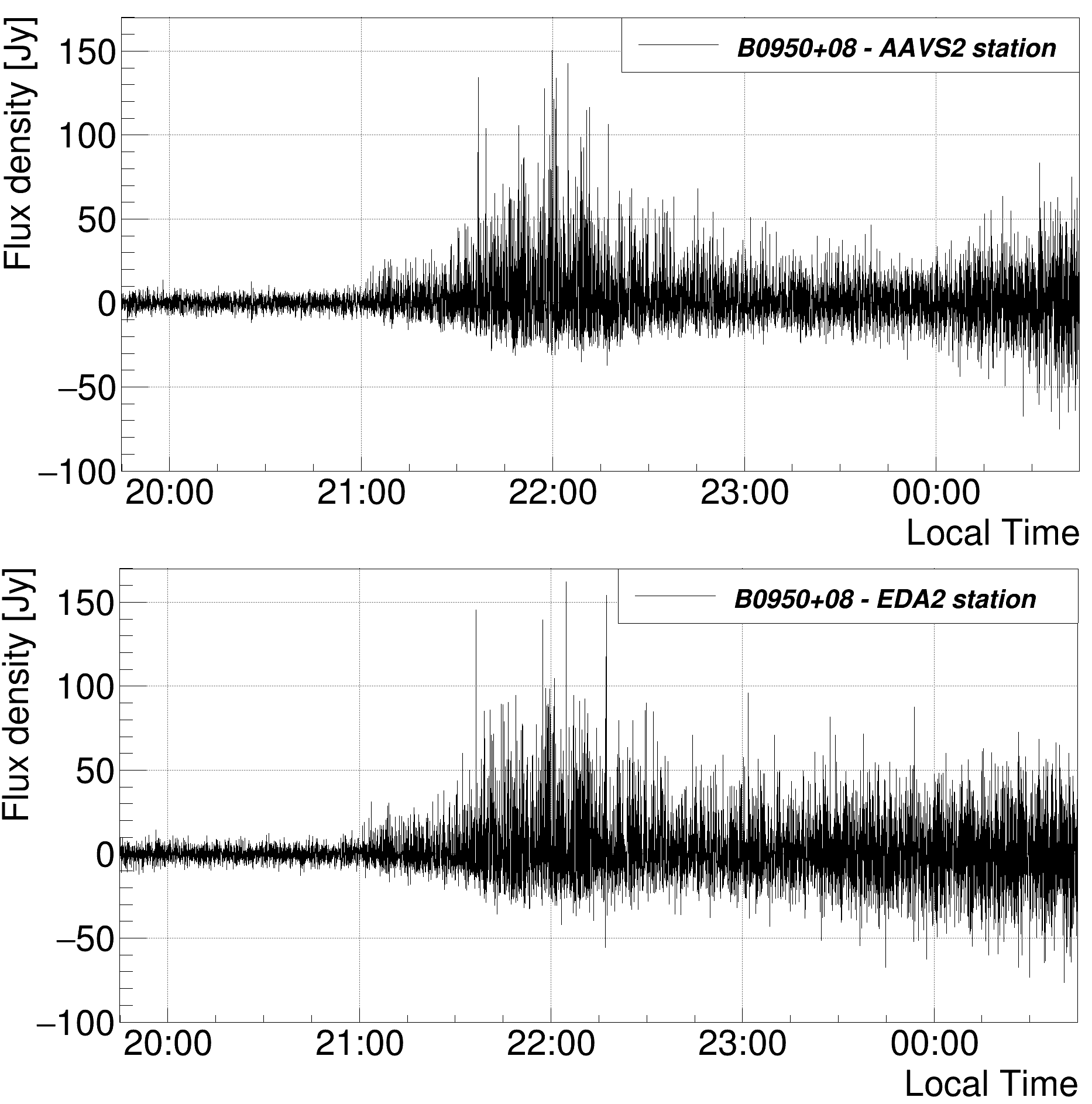}
    \includegraphics[width=\columnwidth]{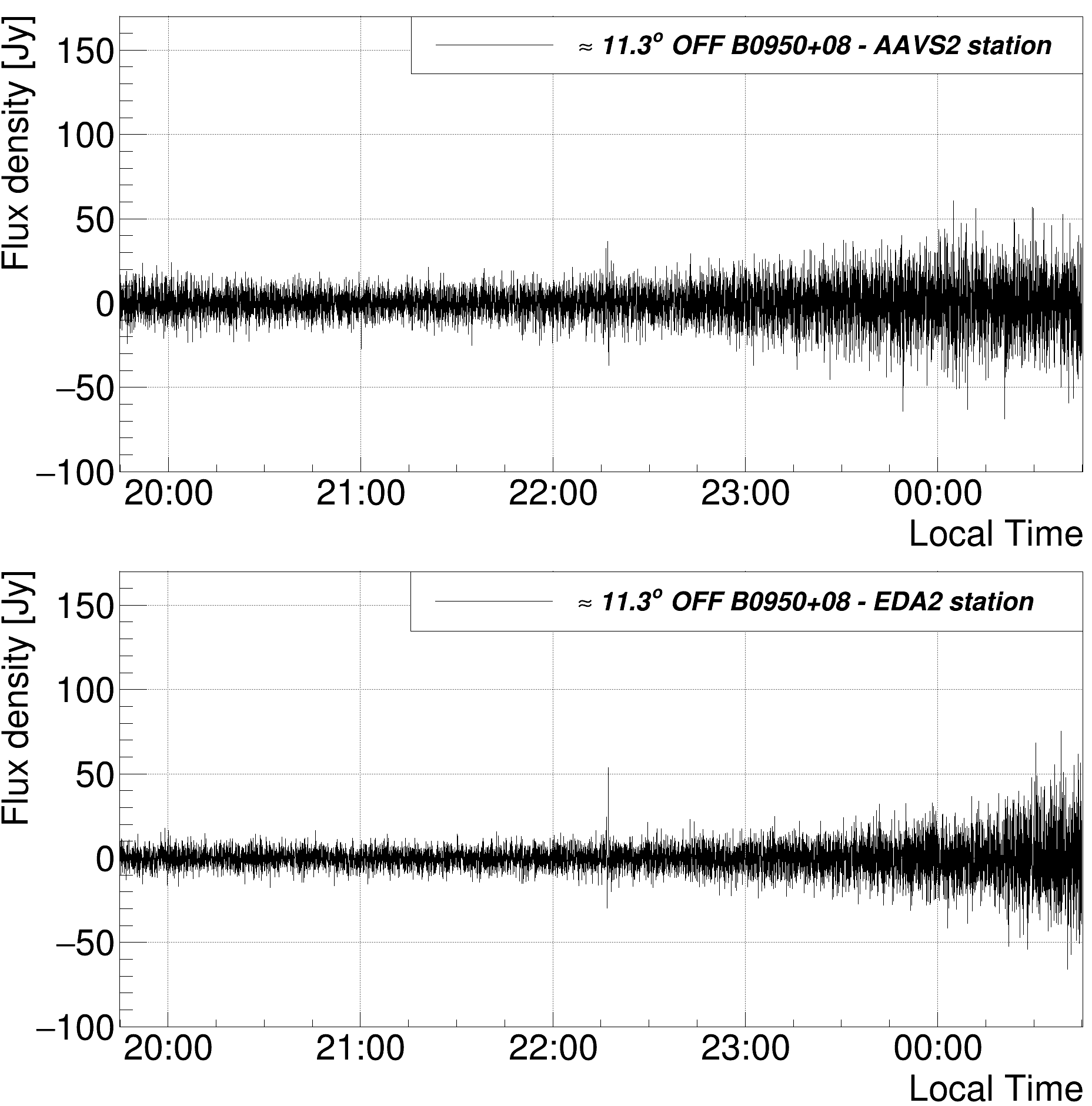}    
	\caption{Left: the lightcurve of PSR B0950+08 based on the data from the EDA2 (lower image) and AAVS2 (top image) collected at 159.375\,MHz between 2020-04-10 11:30 UTC and 2020-04-11 11:30 UTC (only data when pulsar was above elevation $\approx$30\degree\,are shown). It was obtained from the flux density at the position of pulsar and after subtraction of the running median of 30 points around this timestamp (excluding the value at the current timestamp). The flux densities are consistent between both stations and the peak flux density reached maximum of $\approx$150\,Jy/beam (i.e. fluence $\approx$ 300 kJy ms) at 2020-04-10 at 14:04:47 UTC. The standard deviation of the noise in the pulsar ``quiet time'' (before 13 UTC) was approximately 3.8\,Jy/beam and 3.6\,Jy/beam for the EDA2 and AAVS2 stations respectively. Right: a reference lightcurve at a position slightly away ($\approx$11.3\degree) from the pulsar with the EDA2 (lower image) and AAVS2 (top image). The standard deviation of the noise is approximately 4.9\,Jy/beam and 7.2\,Jy/beam for the EDA2 and AAVS2 respectively (for the data before 13:00 UTC). In both lightcurves the noise increased at later times after the Galactic Centre rose above the horizon at 13:30 UTC and especially after it reached 30\degree\, elevation at around 16:00 UTC.}
    \label{fig_b0950_lc}
 \end{figure*}

Example difference images from the night 2020-04-10/11 with a running median of 30 images subtracted are shown in Figure~\ref{fig_b0950_examples}.
The detection of  extreme activity from the pulsar PSR B0950+08, with maximum Stokes I flux density reaching 150\,Jy/beam, was one of the most intriguing detections so far.
The lightcurves of the pulsar during the night 2020-04-10/11 are shown in the left panel of Figure~\ref{fig_b0950_lc}.
The EDA2 and AAVS2 station beams at 160\,MHz are of order 3\degree. Therefore, a single pixel may contain multiple sources, and in order to show variations at the position of PSR B0950+08 a running median of 30 points before and after each timestamp (excluding the timestamp itself) was subtracted from the flux density values at any given timestamp. Moreover, a reference lightcurve of a neighbouring off-the-pulsar position ($\approx$11.3\degree away from the pulsar) is also shown in the right panel of Figure~\ref{fig_b0950_lc} and except for one RFI spike (at around 22:15 AWST) it does not show any significant flux density variations.
These two lightcurves were constructed from the final Stokes I images after subtraction of the running median. It was verified that, unlike some of the spikes spatially coinciding with the pulsar position caused by RFI, the genuine bright pulses from PSR B0950+08 were visible in images in both X and Y polarisations.

The 2020-04-10/11 data from both stations observing at $\approx$159.4\,MHz show very unusual, extremely bright pulses from the PSR B0950+08 with 278 and 208 pulses from the EDA2 and AAVS2 stations respectively. The brightest observed transients ($\sim$155\,Jy) exceed the mean flux density $\approx$2.37\,Jy at 150\,MHz \citep{1995MNRAS.273..411L}) by a factor up to even $\sim$65. Interestingly, they were initially discovered in the ``blind'' search performed on the 2020-04-10/11 dataset. A very similar ``blind'' detection of extremely bright pulses from PSR B0950+08 was also obtained by the AARTFAAC experiment \citep{2020MNRAS.497..846K}, where the authors concluded that these are micro-second giant pulses (GPs) similar to those observed in the Crab pulsar \citep[PSR B0531+21 or J0534+2200;][]{1968Sci...162.1481S}.

\subsubsection{Verification of activity during multiple nights}
\label{sec_b0950_cleaning}

In an effort to verify how frequent are the episodes of such extreme activity, we have analysed \NBnights nights ($\approx$ \NBtime hours) spread over an interval of nearly 6 months. For a quick assessment of whether the pulsar was active we used difference images and for each station generated a lightcurve using these images with flux density measured at the position of the pulsar over the full interval of each observation. This procedure was prone to occasional bright RFI transients due to reflections or transmissions from satellites or planes. We have also found that relatively bright false detections can be generated by side-lobes from very bright RFI with flux densities of the order of thousands Jy/beam. Therefore, in order to calculate the number of 5 and 10 $\sigma_n$ pulses in the lightcurves, we filtered them by requiring that: (i) there is no spike of the same significance in the lightcurve generated from the pixel values at the off-pulsar position $\approx$11.3$\degree$ away from PSR B0950+08 (ii) there is no very bright RFI (exceeding the threshold of 2500\,Jy/beam) identified in the images from a corresponding station within time interval $\pm$ integration time (2\,s) around the PSR B0950+08 pulse time. The same procedure was uniformly applied to all analysed data to calculate the number of bright PSR B0950+08 pulses in the data from both stations and the results are summarised in Table~\ref{tab_b0950}. 

Besides the 2020-04-10/11, bright pulses were found only in two other datasets. Much fewer (20 by EDA2 and 7 by AAVS2) bright pulses ($\sim$100\,Jy/beam) from PSR B0950+08 were observed in the data from 2020-06-26/28 when both stations also observed at $\approx$159.4\,MHz. Finally, bright pulses from PSR B0950+08 were also detected at another frequency in a single dataset (2020-04-16) when both stations observed at 320.3\,MHz and 11 weak pulses ($\lesssim$24\,Jy/beam) were detected only by the AAVS2 station (Tab.~\ref{tab_b0950}). No sufficiently bright pulses from PSR B0950+08 were detected in other datasets. Hence, based on the total time of $\approx$\NBtime hours on PSR B0950+08 spread over nearly 6 monhts, it is clear that the extreme activity observed in the data 2020-04-10/11 is very rare and appears to be less common than refractive or diffractive scintillation events.

\begin{table*}
\caption{Number of 5 and 10 $\sigma_n$ pulses ($\text{N}_{\text{B0950}}^{5\sigma}$ and $\text{N}_{\text{B0950}}^{10\sigma}$ columns respectively ) from PSR B0950+08 observed in each dataset. These values were calculated using the lighcurves generated from the difference images with additional cleaning criteria (Sec.~\ref{sec_b0950_cleaning}) and used as an indicator of the pulsars's activity, while the lightcurve with the more exact background subtraction (using the running median) was generated only when the pulsar was found to be active.}
\centering
\begin{tabular}{@{}cccccc@{}}
\hline\hline
\textbf{Start Time} & \textbf{Frequencies}$^a$    & \textbf{Observing}$^b$ & $\textbf{N}_{\textbf{B0950}}^{5\sigma}$  & $\textbf{N}_{\textbf{B0950}}^{10\sigma}$ & \textbf{Peak} \textbf{Flux Density} \\
  \textbf{(UTC)}    &     \textbf{(MHz)}          &  \textbf{interval} &        \textbf{EDA2 / AAVS2}                  & \textbf{EDA2 / AAVS2} & \textbf{EDA2 / AAVS2} \\
           &                    &  \textbf{EDA2 / AAVS2} &                                  & & \textbf{(Jy/beam)} \\
           &                    &  \textbf{(hours)}  &                                      & &  \\
\hline%
 2020-04-10 & 159.4 / 159.4 & 5.89    & 278 / 208 & 55 / 43 & 150.0 / 162.0 \\ 
 2020-04-11 & 229.7 / 229.7 & 12.06   & 0 / 0     & 0 / 0   & - \\
 2020-04-16 & 320.3 / 320.3 & 9.23    & 0 / 11    & 0 / 0   & 0 / 24.0 \\
 2020-04-29 & 159.4 / 159.4 & 5.36   & 0 / 0     & 0 / 0   & - \\
 2020-05-05 & 229.7 / -     & 1.24 / -    &  0 / - & 0 / -  & - \\ 
 2020-05-07 & 229.7 / -     & 13.25 / -  &  0 / - & 0 / -  & -  \\ 
 2020-05-08 & 229.7 / -     & 13.05 / -  &  0 / - & 0 / -  & -  \\ 
 2020-05-10 & - / 159.4 & - / 3.24 &  - / 0 & - / 0  & -  \\ 
 2020-05-11 & 229.7 / - &  11.14 / -  & 1$^c$ / - & 0 / -  & - \\ 
 2020-05-16 & 320.3 / - &  24.70 / -  &  0 / - & 0 / -  & -  \\ 
 2020-05-18 & 159.4 / -     &  9.31 / - &  0 / - & 0 / -  & -  \\ 
 2020-05-30 & 159.4 / 229.7 & 11.48 &  0 / - & 0 / -  & -  \\ 
 2020-06-01 & 159.4 / 229.7 &  7.73 &  0 / - & 0 / -  & -  \\ 
 2020-06-26 & 159.4 / 159.4 &  46.2  & 20 / 7 & 0 / 0  & 100.0 / 86.0 \\ 
 2020-07-07 & 159.4 / 229.7 &  9.0 / 12.6  & 0 / 0     & 0 / 0   & - \\ 
 2020-07-09 & 159.4 / 229.7 &  0.00  & 0 / 0 & 0 / 0 & - \\ 
 2020-08-27 &  -    / 159.4 &  - / 3.42  &  - / 0 & - / 0  & -  \\ 
 2020-08-28 & 159.4 /   -   & 39.51 / -  & 0 / - & 0 / -  & -  \\ 
 2020-09-11 & 159.4 / 229.7 &  1.63 / 34.81   & 0 / 0     & 0 / 0   & - \\ 
 2020-09-14 & 159.4 / 229.7 &  22.83 / 4.02  & 0 / 0  & 0 / 0   & - \\ 
 2020-09-18 & 159.4 / 229.7$^d$ &  18.43  & 0 / 0  & 0 / 0   & - \\ 
 2020-09-25 & 159.4 / 229.7 &  12.90   & 0 / 0     & 0 / 0   & -  \\ 
 2020-09-27 & 159.4 / 312.5 &  5.91 / 22.26  & 0 / 0  & 0 / 0   & -  \\ 
 2020-10-01 & 159.4 / 312.5 &  18.43   & 0 / 0     & 0 / 0   & -  \\ 
\hline\hline
\end{tabular}
\newline
\begin{flushleft}
\tabnote{$^a$ Frequencies are approximated to a first decimal digit with the exact frequencies 159.375, 229.6875, 312.5 and 320.3125\,MHz.}
\tabnote{$^b$ When PSR B0950+08 was above elevation 20\degree}
\tabnote{$^c$ This pulse 27.3\,Jy was only observed in images from the X polarisation (and not in Y polarisation). Thus, it was excised as RFI.}
\tabnote{$^d$ Strong RFI was observed during some part of the night at the AAVS2 frequency of 229.7\,MHz, which was possibly caused by a tropospheric ducting event}
\end{flushleft}
\label{tab_b0950}
\end{table*}

\subsubsection{Pulse fluence distribution}
\label{sec_b0950_fluence_distribution}

\begin{figure}[h!]
    \includegraphics[width=\columnwidth]{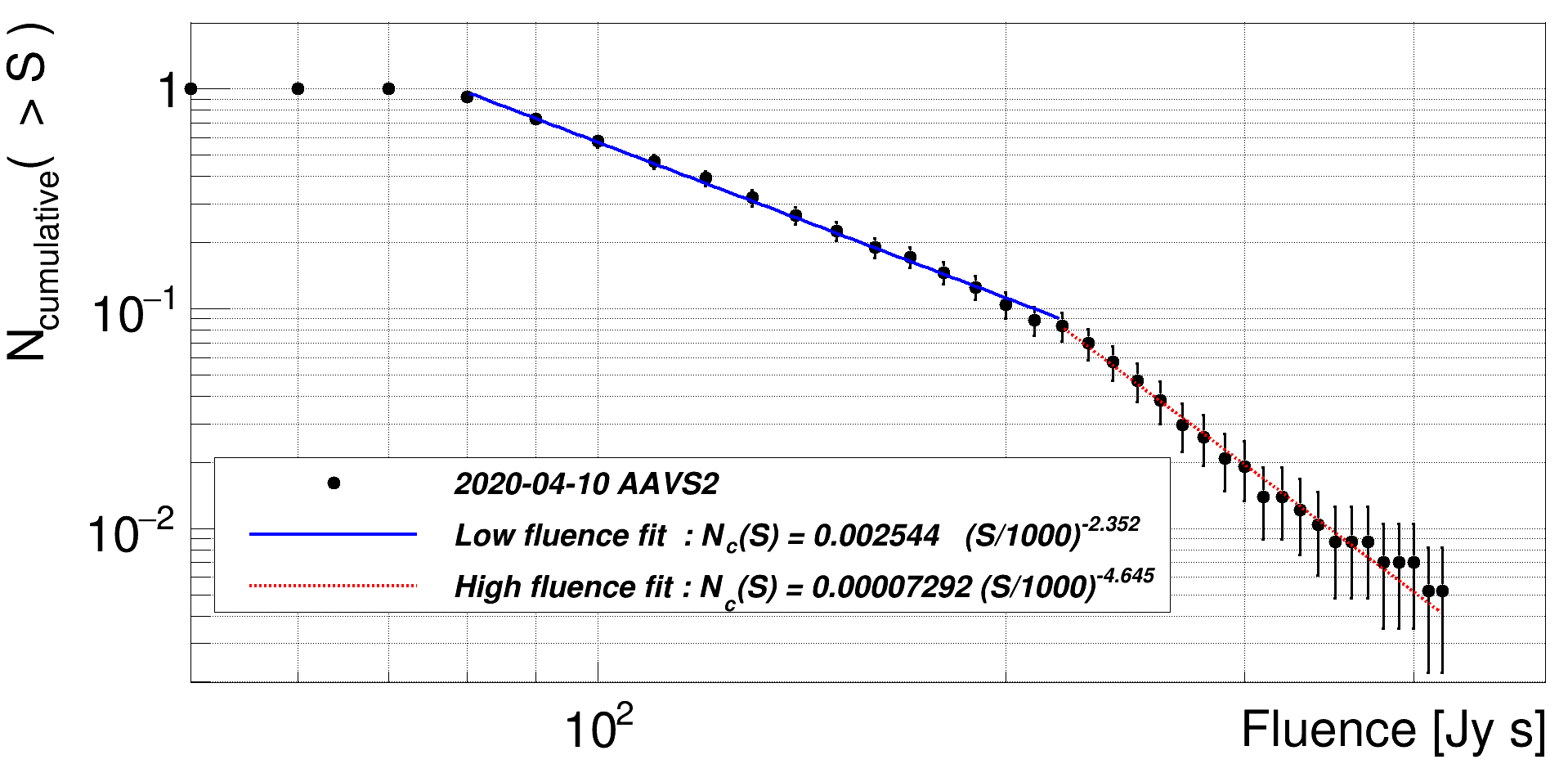}
	\caption{The cumulative distribution of bright pulses from PSR B0950+08 as observed by the AAVS2 station in the 2020-04-10/11 data. The pulse brightness was obtained by subtracting the running median. The distribution can be described with a shallower power law (fitted index $\alpha_{low} \approx -2.4$) below $F_b$=220 Jy s (corresponding to $\approx$360 average pulse fluences) and steeper power law above $F_b$ with $\alpha_{high} \approx -4.6$.}
    \label{fig_b0950_fluence_distr}
\end{figure}

Figure~\ref{fig_b0950_fluence_distr} shows the fluence distribution of the pulses detected with the AAVS2 station obtained after subtracting a running median. The corresponding distribution from the EDA2 station is nearly the same and was not shown for brevity. It can be clearly seen that the slope of the distribution changes at approximate fluence $F_b$=220 Jy s, with the fitted power law index below this $F_b$ value being shallower $\alpha_{low} \approx -2.4$ and steeper above $F_b$ with $\alpha_{high} \approx -4.6$.  This value of $F_b$ corresponds to approximately 360 average pulses (AP) assuming mean flux of 2.37\,Jy at 150\,MHz \citep{1995MNRAS.273..411L} and pulsar period $P=0.253$\,s according to the ATNF pulsar catalogue \citep{2005AJ....129.1993M}. This is very similar to the distributions of GP fluences previously reported by \citet{2020MNRAS.497..846K} and shown in Figure 6 in their paper, where their fitted power law indexes were -2.5 and -4.3 at frequency 58.3\,MHz and -1.9 and -6.8 at 61.8\,MHz. In their work the ``break fluence'' values were $\approx$45 AP fluences at 58.3\,MHz and $\approx$88 AP fluences at 61.8\,MHz.  Moreover, a similarly steep distribution of fluence was also reported by \citet{2016AJ....151...28T} based on the LWA observations at 42 and 74\,MHz with power law indexes -4.09 and -5.06 respectively. On the other hand, \citet{2012AJ....144..155S} reported much shallower (power law index $\approx$ -2.2) GP fluence distribution at 103\,MHz. These previous studies assign the high fluences and the steep slope of their cumulative distribution to be due to intrinsic emission mechanisms.

\subsubsection{Plausible physical mechanisms for the observed extreme activity}
\label{sec_b0950_discussion}

The limited bandwidth and time resolution of current observations prevent us from making firm conclusions about the physical mechanisms that may have caused the observed extreme activity. It may be due to intrinsic effects, such as large-amplitude pulses (e.g. giant pulses) that are part of the pulsar emission process, or caused by propagation effects (e.g. refractive or diffractive scintillation). Even though disentangling between the two is difficult, we comment on the related possibilities, on the basis of our observations and analysis. 

As discussed in Section~\ref{sec_b0950_fluence_distribution},  a cumulative fluence distribution of the observed bright events (pulses) suggests a break in the power-law index, which hints there being two populations of pulses.  The brighter population exhibits a substantially larger power-law index of $-4.6$, akin to those observed for giant pulses from the Crab-like pulsars \citep{2007A&A...470.1003P,2008ApJ...676.1200B,2013MNRAS.430.2815Z,2019MNRAS.483.4784M}. Propagation effects may have enhanced their apparent flux densities, but is unlikely to cause such a break in the cumulative distribution. 

As noted earlier, our measured flux densities of pulses (up to $\sim150$\,Jy/beam) are significantly larger than the mean flux densities expected for this pulsar at 150 MHz ( $\approx$2.37\,Jy, cf.  \citet{1995MNRAS.273..411L}). The observed flux density increase (by a factor of $\sim$65), is well over order of magnitude more than typical  amplifications that can be attributed to refractive scintillation, which tends to enhance the flux density by a factor of ~2-3. Indeed there is observational evidence that nearby pulsars tend to exhibit much higher variability e.g.  flux density modulations of $\sim$5 -- 6 have been reported for PSR B0950+08 \citep{2016MNRAS.461..908B} and PSR J0437-4715 \citep{2016MNRAS.461..908B,2014ApJ...791L..32B}. The measured boost in flux density is thus almost an order of magnitude larger than that we may have expected from refractive scintillation alone. The relative rarity of  occurrence (3 nights out 24 spread over nearly 6 months) further supports such a conjecture. A similar possibility was also suggested by the recent work of \citet{2020MNRAS.497..846K}.

However, we note that MWA observations of this pulsar (\citet{2016MNRAS.461..908B}) have reported peak flux densities of  $\sim$48.6\,Jy near these frequencies, which they attribute to diffractive scintillation. The fact that our measured flux densities are only $\sim$3 times larger may suggest a similar possibility, particularly considering their relatively coarser time resolution (112\,s images), due to which a significantly larger number of pulses are averaged (i.e. $\sim$448 vs. 8). In other words, it is possible that individual pulses in their observations may have been as bright as those seen in our observations. However, a major distinction is, in our case, that the observed activity tends to last on significantly longer timescales ($\sim$ 2\,hours) than the expected timescale for diffractive scintillation, which is $\sim$30 min, based on the analysis of \citet{2016MNRAS.461..908B}. Therefore, it is possible that the extreme activity seen in our data is likely a more complex form of propagation effects, e.g. flux density boosting caused by both the effects, and our observations capturing a bright scintle near its diffractive scintillation peak, when the pulsar's mean flux  density was near its refractive scintillation peak at the time of observation. While the timescale and bandwidth of \citet{2016MNRAS.461..908B} are consistent with the expectations based on earlier low-frequency observations of the pulsar \citet{1992Natur.360..137P}, we note that our observations suggest the activity lasting for a longer period of time ($\sim$ 2\,hours). A large amplification from diffractive scintillation has also been reported for another nearby pulsar PSR B0655+64 by \citet{1997A&A...325..631G}, who observed $\sim$43 factor amplification in their data, and disfavoured the intrinsic effects in favour of diffractive scintillation as an explanation of such a large increase in brightness of this specific pulsar.

In short, while it is possible that the observed flux density enhancement can in principle be attributed to diffractive scintillation, the timescale of the activity and the indication of a break in the slope of cumulative fluence distribution are suggestive of somewhat different mechanisms. Future observations at higher time resolutions  and over a larger bandwidth may help to resolve this.

\subsection{Other astrophysical candidates}
\label{sec:other_astro_candidates}

As shown in Table~\ref{tab_candidates}, besides transients from PSR B0950+08 and RFI, several transients of astrophysical origin were detected in the data. One of them was observed as multiple pulses from the same position (other than PSR B0950+08) in the sky and is very similar to the initial ``blind'' detection of the pulsar PSR B0950+08 in the 2020-04-10/11 data. This is a potentially interesting astrophysical object and is currently being investigated.

\red{Our transient searches were not optimised for longer timescales. Nevertheless, we note that, all the identified candidates were seen in single 2\,s images only (images after the detection difference images were verified not to have any signals). Hence, we have not detected events visible in more than one subsequent 2\,s images. Neither, we observed prolonged periods of activity with different sources at some parts of the sky being magnified by ionospheric magnification events on timescales $\gtrsim$10\,s as reported by \citet{2020arXiv200311138K} using AARTFAAC system.  The most likely explanation for the lack of such detections are our higher observing frequencies than the AARTFAAC observations at 60\,MHz, which based on the equation 5 in their paper reduces the potential magnifications by a factor $\sim(60.00/160)^2\sim$0.14. However, we expect to observe similar events in the future when we start observations and transient searches (including longer timescales) at low frequencies (especially $\le$100\,MHz), which we have not tried yet. We also note that our detections of activity from PSR B0950+08 and the other similar object (during different nights) were limited to a single active source per night, lasted several hours and were observed only on timescales $\lesssim$2\,s. Hence, they were unlikely caused by similar effects.}

\section{LIMITS ON LOW-FREQUENCY EMISSION FROM FRBs}
\label{subsec:frb_limits}

The monitoring system was routinely used since April 2020 whenever stations were observing in a standalone interferometer mode. 
On 2020-09-14 and 2020-09-19, EDA2 and AAVS2 were observing at 159.375 and 229.6875\,MHz when they serendipitously co-observed FRBs 200914 and 200919 \citep[ATel \#14040;][]{2020ATel14040}, which were detected by the Deeper Wider Faster \citep[DWF;][]{2019IAUS..339..135A} program using Parkes Radio-Telescope. The 1-$\sigma$ upper limits, of the order of $25 - 33$ kJy ms, were derived from 2\,s images using difference images and reported in \citep[ATel \#14044;][]{2020ATel14044}. These limits can be scaled to predict that with the millisecond time resolution of images the 1-$\sigma$ limits will be of the order of 670 and 2100 Jy ms for integration time 1 and 10 ms respectively with the original bandwidth of $\approx$0.926\,MHz and assuming an approximate 30 kJy ms limit from 2\,s images. Moreover, increasing the bandwidth to $\sim$50\,MHz will further improve the 1-$\sigma$ limits to $\sim$ 85 and 265 Jy$\cdot$ms for 1\,ms and\,10 ms integration times respectively, which in the light of the recent LOFAR results predicting 3-450 FRBs/sky/day above 50 Jy ms at 90\% confidence level \citep{2020arXiv201208348P}, gives very good prospects for FRB detections with the upgraded back-end systems for the SKA-Low stations.

\section{LIMITS ON RADIO TRANSIENT RATE}
\label{subsec:rate_limits}

Given the limitations of the system in terms of time and frequency resolutions, as well as bandwidth, and the difficulties of excising all the false positives due to RFI from satellites (Section~\ref{subsec:results_satellites}), we cannot unambiguously determine which transients are of genuine astrophysical origin.
This is especially the case when both stations observed at the same frequency. However, as can be seen from Table~\ref{tab_candidates}, it is much easier to excise RFI due to satellites when the two stations observed at different frequencies. Therefore, we used 7 datasets when EDA2 observed at 159.4 MHz and AAVS2 at 229.7 MHz, corresponding to a total observing time $\approx$\sdlimittime\,hours. We did not include the datasets when EDA2 observed at 159.4\,MHz and AAVS2 at $\ge$312.5\,MHz in order to keep the data uniform, at the expense of a small ($\sim$10 -- 20\%) reduction in the total observing time. In the above cases, the only source of broadband RFI that we detect is the satellite BGUSAT, as previously discussed in Section~\ref{subsec:results_satellites}.

Using the 7 dual-frequency 159.4/229.7 MHz datasets, we identified one astrophysical object, which generated multiple transients on 2020-05-30, and is currently being investigated as a potential new pulsar candidate.
Besides these detections, there were no other transient candidates identified in the dual-frequency data. Therefore, assuming that the particular candidate from the 2020-05-30 dataset is a different class of object (repeating and likely a Galactic pulsar) than other reported short-timescale, low-frequency transients \citep{2016MNRAS.456.2321S,2019ApJ...874..151V,2020arXiv200311138K,2020arXiv200313289K}, we derived a preliminary surface density upper limit on non-repeating transients in the frequency range 159.4 -- 229.7 MHz (note: we required transients to be detected at both frequencies), following the procedure outlined by \citet{2016MNRAS.456.2321S} and references therein. 

\subsection{Calculation of a single transient detection threshold}
\label{subsec:transient_density_cutoff_threshold}

The mean transient detection flux density threshold of our study was calculated as 5$\sigma_m$, where $\sigma_m$ is a mean standard deviation (RMS) noise level. Because we were searching for transients in all-sky difference images (elevation $\ge$25\degree), the RMS noise has a strong dependence on the zenith angle, i.e. the distance from the centre of an image, due to the beam correction (required to provide the correct flux density scale). Moreover, the noise level changes with the local sidereal time (LST). Hence, in order to provide a single number for the RMS noise level averaged both spatially and over time, for each all-sky difference image we calculated an average RMS by generating a noise map following the procedure described in Section 5.3.2 in \cite{2020arXiv201208075S}. In brief, for each image pixel, a corresponding local RMS was calculated as the standard deviation of all pixel values in a radius of 10 pixels. Then, the average RMS ($\sigma_i$) was calculated as the mean of all the local RMS values over the entire noise map (i.e. the mean of all pixels at elevation $\ge$25\degree). This averaging of the RMS is effectively equivalent to a procedure used by \citet{2016MNRAS.456.2321S}, where the authors calculated an average RMS noise level over a much smaller field of view (FoV) as an area-weighted average RMS calculated in three rings around the centre of the image. Finally, we excluded bad images ($\sigma_i >$  15\,Jy), and averaged the $\sigma_i$ values over all observed LSTs. This resulted in a single averaged value of the RMS noise level: $\sigma_m \approx$8.5\,Jy (i.e. $5\sigma_m\approx$42\,Jy), which was consistent between both stations and frequencies. 

\subsection{Calculation of a transient surface density upper limit}
\label{subsec:transient_density_limit_calculation}

For this calculation, the original FoV of the sky above elevation 25\degree, $\Omega\approx$\fovdegree\,deg$^2$,  was multiplied by a correction factor $\approx$\fovreduction\, to account for regions of the sky excluded by  criteria 1--4 in Section~\ref{subsec:filtering_transients}. This correction factor was estimated using a Monte Carlo simulation; the corrected FoV  $\Omega_\text{corr}\approx$\fovdegreecorr\,deg$^2$.  The reduction in the total observing time ($\approx$\sdlimittime\,hours) due to strong RFI (criterion 8 in Section~\ref{subsec:filtering_transients}) was negligible ($\approx$0.0062\,hours) when the stations observed at 159.4 (EDA2) and 229.7 (AAVS2) MHz. Given that the number of 2\,s images (epochs) was $N$=\sdlimitepochs, the transient surface density upper limit at the 95\% confidence level was then calculated as $-\text{ln}(0.05)/(\Omega_\text{corr} \times (N-1)) = \surfacedensitylimit \text{deg}^{-2}$ for flux densities brighter than $42$\,Jy (5$\sigma$), and on a timescale of 2\,s for each individual epoch. 

\subsection{Comparison with other low-frequency surveys}
\label{subsec:comparisons_with_others}

 While a detailed analysis is beyond the scope of this paper, the upper limit calculated in Section~\ref{subsec:transient_density_limit_calculation} is \red{at least several times higher (up to a factor $\approx$20)} than the transient surface densities reported by \citet{2015JAI.....450004O}, \citet{2019ApJ...874..151V} and \citet{2020arXiv200313289K}; for example, see Fig. 3 in the latter study, or Fig. 6 in \cite{2019ApJ...886..123A}. Our sensitivity level ($42 \pm 15$\,Jy) is deeper than the flux densities of the transients detected by \citet{2019ApJ...874..151V} and \citet{2020arXiv200311138K,2020arXiv200313289K}, as well as the sensitivity levels in the \citet{2015JAI.....450004O} study, with the caveat that our study was conducted at higher observing frequencies. If the underlying transient population has a steep spectral index, then our effective sensitivity level at lower frequencies is more directly comparable to those achieved in the aforementioned studies, being a similar order of magnitude. \red{Our upper limit on the transient surface density adds a new measurement at short timescales: between the limits at 5\,s reported by \citet{2015JAI.....450004O} and the shortest reported timescale of 1\,s in \cite{2020arXiv200313289K}. This a relatively unexplored region of parameter space.}

\section{OTHER APPLICATIONS}
\label{subsec:other_aplications}

The presented all-sky imaging system has been routinely used for real-time monitoring of the EDA2 and AAVS2 data acquisitions and quick assessment of the data quality for the datasets collected since the end of 2019. As described in AAVS1 description paper (Bentham et al. submitted) difference imaging technique is a very convenient way of calculating station sensitivity expressed as System Equivalent Flux Density (SEFD) at zenith. Real-time calculation of SEFD and comparison with the sensitivity predicted by the simulations is one of the planned extensions to the system. This will enable real-time monitoring of the SKA-Low stations sensitivity, which is a critical characteristic of the radio-telescope performance.

\section{SUMMARY}
\label{sec:summary}

We presented the first real-time all-sky imaging system in the Southern Hemisphere operating at low radio-frequency implemented on the two prototype stations (EDA2 and AAVS2) of the SKA-Low radio-telescope. The all-sky 2\,s images generated by the system have been searched for radio-transients using transient identification pipeline based on difference imaging. Many long observations (up to even 6 days) were conducted with both stations collecting data in parallel at the same or different frequency channels and the resulting images have been analysed in search for transients. The search algorithm requiring the transient candidate to be detected in corresponding difference images from both stations (coincidence) was executed on the data from \Ncoincnights nights ($\approx$ \Ncoinctime hours) when both stations were collecting data simultaneously. 

The majority of detected transients are due to RFI emissions or reflections from the satellites, aircraft or meteors as RFI from ground-based transmitters were excised by imposing elevation $\ge$25\degree \, criteria. However, even with the existing limitations of the system, a small number of interesting transients of confirmed astrophysical origin have been identified. The most interesting detections of astrophysical phenomena were extremely bright pulses from the pulsar PSR B0950+08. Similar activity was reported by the AARTFAAC experiment and claimed them to be due to giant pulses similar to those produced by the Crab pulsar \citep{2020MNRAS.497..846K}. The highest activity of the pulsar was observed during 3\,hours of the night 2020-04-10/11 when pulses as bright as 150\,Jy/beam (fluence 300\,Jy s/beam) were detected in 2\,s images with over 208 and 278 pulses exceeding 5$\sigma_n$ thresholds of 21\,Jy/beam and 18\,Jy/beam for AAVS2 and EDA2 respectively. In total, data from \NBnights nights ($\approx$ \NBtime hours) spread over nearly 6 months were used to create ligthcurves at the position of PSR B0950+08, but pulses were detected only in 3 of them. Besides the original detections in the 2020-04-10/11 dataset, in the 2020-06-26/27 data $\sim$10-20 (ten times less) pulses from PSR B0950+08 up to $\sim$100\,Jy/beam were also detected by the both stations, whilst in the 2020-04-16/17 data 11 weak pulses were detected by AAVS2 only (both stations observed at 320.3\,MHz). The extreme brightness of these events, steep slope (fitted power law index $\approx$-4.6) of the cumulative fluence distribution of the brighter population of transients (fluence $\gtrsim$\,220~Jy~$\cdot$~s), and rare occurrence (only 3 out of 24 nights spread over nearly 6 monhts) makes it unlikely to be entirely due to diffractive scintillation and indicates another mechanisms, such as combination of diffractive and refractive scintillation or intrinsic emission mechanisms (e.g. giant pulses as suggested by earlier study by \citet{2020MNRAS.497..846K}). However, we leave further analysis and conclusions to the future publications. We have also detected an unknown astrophysical object showing similar bright pulses over about 1\,hour on two subsequent days, which is currently being investigated. 

The small observing bandwidth ($\approx$0.926\,MHz), which is the main limitation of the current system, prevented us from analysing spectral properties of the bright pulses from PSR B0950+08. However, these detections clearly demonstrate that very bright pulsars (even new) and/or other transient objects, such as FRBs can be successfully detected with this system and potentially trigger more sensitive instruments, such as the MWA. Especially, once the system is upgraded with more observing bandwidth (of the order of 50\,MHz) and millisecond time resolution. 

\red{Finally, using the observations at different frequencies (EDA2 at 159.4\,MHz and AAVS2 at 229.7\.MHz), we derived a preliminary transient surface density upper limit of $\surfacedensitylimit \text{deg}^{-2}$ for a timescale of 2\,s and a $5\sigma$ sensitivity level of $42 \pm 15$\,Jy. While our upper limit is not as constraining as other results from previous low-frequency studies in the literature, this is one of the shortest timescales for which a surface density, or surface density upper limit, has been reported thus far.}

The system can also be used for other purposes, such as continuous monitoring of SKA-Low stations sensitivity and RFI studies at the MRO in real-time. Particularly, with better automatic classification of identified events it will be possible to continuously catalogue all the RFI detections to be later used in science data analysis.

\section{FUTURE PLANS}
\label{sec:future}

In order to increase the sensitivity to short pulses, we are planning to upgrade the system with more instantaneous bandwidth (of the order of 50\,MHz) and millisecond time resolution. Furthermore, the data will also be fine channelised and the spectral information will further help with classification and distinguishing between different types of events and RFI excision in particular. These improvements will enable real-time image-based searches for dispersed radio pulses such as FRBs. We estimate that at least 50-100 FRBs per year can be detected by such an extended system (assuming continuous operation). 

Moreover, the system will also be enhanced with the triggering capability to react to external FRB triggers from ASKAP CRAFT \citep{2010PASA...27..272M}, UTMOST \citep{2017PASA...34...45B} or Parkes Radio telescope.
In addition, we will react to alerts from transients distribution networks, such as Gamma-ray Bursts Coordinate Network (GCN)\footnote{https://gcn.gsfc.nasa.gov/}, VO-Events \citep{2014A&C.....7...12S,2016arXiv160603735S} etc. Therefore, upon receiving the trigger complex voltages from all antennas will be recorded in full time resolution before, during and after the burst detection by high frequency instruments, which will be enabled by a voltage buffer. However, even before the upgrade to higher time resolution and wider bandwidth, we will enable automatic formation of station beam in the direction of the externally provided transient coordinates and record station beam complex voltages for off-line analysis.

Besides, these major developments we are also planning several smaller software improvements. The system will automatically perform calibration using data collected during Sun transits. Several existing source finders will be tested to select the most optimal for the SKA-Low station all-sky difference images. Further, we will test the posibility of using real-time plane tracking systems to automatically excise RFI caused by these objects. Using wider observing bandwidth and spectral information from fine channalisation of complex voltages, we will improve excision of candidates caused the RFI transmissions and/or reflections. This will also enable real-time cataloguing of RFI events to a database, which will be extremely valuable for developing future observing strategies with the SKA-Low. We will also start using full polarimetric information and form Stokes images (I, Q, U and V), which will help with RFI excision (known to be polarised) and potentially enable idendification of pulsar candidates in Stokes V images. With the improvements in the RFI excision we will start testing automatic algorithms for classification of identified transient candidates in order to further reduce the number of events which require visual inspection. Finally, as discussed in Section~\ref{subsec:other_aplications}, the system will be extended with real-time measurements of station sensitivity and comparisons against the expectations based on simulations.

\begin{acknowledgements}
AAVS2 and EDA2 are hosted by the MWA under an agreement via the MWA External Instruments Policy.
This scientific work makes use of the Murchison Radio-astronomy Observatory, operated by CSIRO. We acknowledge the Wajarri Yamatji people as the traditional owners of the Observatory site. This work was further supported by resources provided by the Pawsey Supercomputing Centre with funding from the Australian Government and the Government of Western Australia. The acquisition system was designed and purchased by INAF/Oxford University and the RX chain was design by INAF, as part of the SKA design and prototyping program.

We acknowledge the work and support of the developers of the following following Python packages: Astropy  \citep{astropy:2013,astropy:2018} and Numpy \citep{2011CSE....13b..22V}. This research has made use of NASA's Astrophysics Data System. 

\end{acknowledgements}

\bibliographystyle{pasa-mnras}
\bibliography{eda2tv}

\end{document}